\RequirePackage{lineno} 
\documentclass[prd, reprint, preprintnumbers, twocolumn, eqsecnum,floatfix,letterpaper,superscriptaddress,nofootinbib]{revtex4-1}
\usepackage{graphicx,url,amssymb,amsmath,rotating,color,units,wasysym,epsfig,multirow,epstopdf,subcaption,xcolor}
\usepackage[normalem]{ulem}
\usepackage{comment}
\usepackage[multiple]{footmisc}
\usepackage[colorlinks,urlcolor=blue,citecolor=blue,linkcolor=blue]{hyperref}
\usepackage{cleveref}
\usepackage{multirow}
\usepackage{graphicx}
\usepackage{lineno}
\usepackage{arydshln}

\graphicspath{{./}{figures/}}

\newcommand{\red}[1]{{\color{red} (#1)}}

\newcommand{\Ethan}[1]{{\color{blue} (Ethan: #1)}}

\begin{document}

\title{Likelihood-free inference for gravitational-wave data analysis and public alerts}
\author{Ethan Marx}
\affiliation{Department of Physics, MIT, Cambridge, MA 02139, USA}
\affiliation{LIGO Laboratory, 185 Albany St, MIT, Cambridge, MA 02139, USA}
\author{Deep Chatterjee}
\affiliation{Department of Physics, MIT, Cambridge, MA 02139, USA}
\affiliation{LIGO Laboratory, 185 Albany St, MIT, Cambridge, MA 02139, USA}
\author{Malina Desai}
\affiliation{Department of Physics, MIT, Cambridge, MA 02139, USA}
\affiliation{LIGO Laboratory, 185 Albany St, MIT, Cambridge, MA 02139, USA}

\author{Ravi Kumar}
\affiliation{Department of Aerospace Engineering, IIT Bombay, Powai, Mumbai, 400076, India}
\affiliation{School of Physics and Astronomy, University of Minnesota, Minneapolis, MN 55455, USA}

\author{William Benoit}
\affiliation{School of Physics and Astronomy, University of Minnesota, Minneapolis, MN 55455, USA}

\author{Argyro Sasli}
\affiliation{School of Physics and Astronomy, University of Minnesota, Minneapolis, MN 55455, USA}

\author{Leo Singer}
\affiliation{Astroparticle Physics Laboratory, NASA Goddard Space Flight Center, Code 661, Greenbelt, MD 20771, USA}

\author{Michael W. Coughlin}
\affiliation{School of Physics and Astronomy, University of Minnesota, Minneapolis, MN 55455, USA}
\author{Philip Harris}
\affiliation{Department of Physics, MIT, Cambridge, MA 02139, USA}
\author{Erik Katsavounidis}
\affiliation{Department of Physics, MIT, Cambridge, MA 02139, USA}
\affiliation{LIGO Laboratory, 185 Albany St, MIT, Cambridge, MA 02139, USA}

\date{\today}


\begin{abstract}
Rapid and reliable detection and dissemination of source parameter estimation data products from gravitational-wave events, especially sky localization, is critical for maximizing the potential of multi-messenger astronomy. Machine learning based detection and parameter estimation algorithms are emerging as production ready alternatives to traditional approaches. Here, we report validation studies of AMPLFI, a likelihood-free inference solution to low-latency parameter estimation of binary black holes. We use simulated signals added into data from the LIGO-Virgo-KAGRA's (LVK's) third observing run (O3) to compare sky localization performance with BAYESTAR, the algorithm currently in production for rapid sky localization of candidates from matched-filter pipelines. We demonstrate sky localization performance, measured by searched area and volume, to be equivalent with BAYESTAR. We show accurate reconstruction of source parameters with uncertainties for use distributing low-latency coarse-grained chirp mass information. In addition, we analyze several candidate events reported by the LVK in the third gravitational-wave transient catalog (GWTC-3) and show consistency with the LVK's analysis. Altogether, we demonstrate AMPLFI's ability to produce data products for low-latency public alerts.

\end{abstract}
 \maketitle

\section{Introduction}  \label{sec:intro}
A decade has passed since the first direct detection of gravitational waves (GWs) \cite{PhysRevLett.116.061102}. Now, observing GWs from compact binary coalescences (CBCs) is commonplace. The number of GW candidates has increased by two orders of magnitude in this decade, rising from three candidates in the first observing run, to over three-hundred cumulative after the fourth observing run (O4) \cite{AbEA2018b, theligoscientificcollaboration2022gwtc21, Abbott_2023,LIGOScientific:2025slb}. The discovery rate is expected to continue to rise as the Advanced LIGO~\cite{Aasi_2015}, Advanced Virgo~\cite{VIRGO:2014yos} and KAGRA~\cite{kagra} interferometers approach the design sensitivity era of the fifth observing run \cite{ observing_scenarios}. 
Accompanying this rapid growth has been excitement to transmit real-time public alerts to the broader astrophysical community, with the goal of finding coincident electromagnetic (EM) signals. 

Machine learning (ML) methods for detecting and characterizing CBCs are emerging as production ready solutions to real-time GW data analysis. For detection, many methods are being developed, showing promise in achieving the sensitivity of traditional matched-filter pipelines at a reduced latency and computational cost \cite{aframe-methods, sage, Nousi_2023}. However, these methods do not provide a solution for producing the suite of low-latency data products necessary for informing effective EM followup after an initial detection. Currently, the LVK's low-latency alert infrastructure relies on data products expected from matched-filter pipelines. For example, the production sky localization algorithm for events from matched-filter pipelines, BAYESTAR, requires signal to noise ratio (SNR) time series \cite{bayestar}. In addition, coarse-grained chirp mass information (a recent addition to LVK alerts\footnote{\url{https://emfollow.docs.ligo.org/userguide/}}) and other source property classifiers utilize point estimates from the best fit matched-filtering template \cite{Chatterjee:2019avs}. ML-based search algorithms do not naturally provide SNR time series or template parameters required by these downstream tasks. 

In Ref.~\cite{Chatterjee:2024pbj}, we introduced AMPLFI\footnote{\textbf{A}ccelerated \textbf{M}ultimessenger \textbf{P}arameter estimation using \textbf{LFI}; pronounced ‘amp-li-fy’.}, a parameter estimation (PE) algorithm based on likelihood-free inference (LFI), as a tool for providing low-latency PE data products. We motivated AMPLFI as a real-time followup to ML-based detection algorithms, specifically, Aframe \cite{aframe-methods}. In this paper, we describe recent improvements to AMPLFI, and demonstrate its ability to produce robust low-latency source PE data products. 

This paper is organized as follows. In Sec.~\ref{sec:lfi} we provide a summary of LFI in GW astronomy. In Sec.~\ref{sec:review} we discuss the AMPLFI algorithm previously presented in Ref.~\cite{Chatterjee:2024pbj} including improvements made since. In Sec.~\ref{sec:testing-data} we present the dataset used to evaluate AMPLFI's performance. In Sec.~\ref{sec:sky-localization}, we discuss and benchmark our method for producing sky localizations from AMPLFI posterior samples. In Sec.~\ref{sec:results-skyloc} we compare AMPLFI's sky localization performance with BAYESTAR \cite{bayestar}. In Sec.~\ref{sec:results-consistency} we demonstrate the self-consistency of AMPLFI sky localizations. In Sec.~\ref{sec:results-robustness}, we illustrate AMPLFI's robustness to changing detector background over a time scale of several months. Finally, in Sec.~\ref{sec:results-gwtc3} we analyze real GW candidates from GWTC-3 and show consistency with published GWTC-3 results.

\section{Likelihood Free Inference} \label{sec:lfi}
In GW astronomy, estimating the parameters $\theta$ of a source given strain data $d$ is typically done using Bayesian inference. Traditional Bayesian PE algorithms rely on repeated evaluations of an explicit likelihood function, utilizing stochastic sampling techniques such as Markov Chain Monte Carlo (MCMC) or nested sampling to explore the signal parameter space \cite{bilby_paper, lalinference_paper}. For GW PE, this process can be arduous. Each likelihood evaluation requires simulating a waveform, which can be expensive. In addition, the full 15 dimensional signal parameter space must be fully explored. Low-latency PE methods typically make approximations, to either the likelihood function or waveform physics, to simplify the exploration of the posterior \cite{Canizares:2014fya,Morisaki:2023kuq,Morisaki:2020oqk,SiPr2016}.

Posterior estimation using LFI involves training a probabilistic neural network with tuneable parameters $\phi$, $q_\phi(\theta | d)$, to approximate the true posterior $p(\theta | d)$. A type of probabilistic neural network known as a conditional normalizing flow~\cite{papamakarios2021normalizing} is a popular choice for constructing the approximation $q_\phi(\theta | d)$. Conditional normalizing flows consist of flexible, parameterized variable transforms which map between a simple base distribution, $\pi(u)$, typically taken to be a standard normal, and a more complex distribution which is being approximated (e.g. $q_\phi(\theta|d)$). The normalizing flow is conditional in the sense that the parameters that define the transformations from the base distribution to the more complex distribution are a function of the data $d$. Normalizing flows have been utilized in GW astronomy for a variety of use cases \cite{Dax_2021, Dax_2025, Williams:2021qyt, PhysRevD.109.102004}.

Training normalizing flow models requires large quantities of data realizations from the likelihood. This means a diverse set of both noise instances, $n$, and signal simulations $h(\theta)$ such that one can simulate $d = n + h(\theta) \sim p(d|\theta)$. The utility of LFI is that once the approximator is trained, generating samples from the posterior $\theta \sim q_\phi(\theta | d)$ can be done rapidly by drawing samples from the base distribution and applying the learned transformations to obtain the posterior sample. Drawing thousands of samples takes seconds when done on accelerated hardware like GPUs. In addition, LFI also provides density estimation, or scoring, of samples i.e. $p(\theta \vert d)\approx q_\phi(\theta \vert d)$.

\section{AMPLFI} \label{sec:review}
AMPLFI is a PE algorithm based on LFI and was presented in detail in Ref.~\cite{Chatterjee:2024pbj}. In Sec.~\ref{sec:amplfi-summary} we summarize the algorithm and in Sec.~\ref{sec:improvements} we discuss in more detail improvements that have been made regarding training data usage and neural network architecture.

\subsection{Summary} \label{sec:amplfi-summary}
Applications of LFI with normalizing flows to GW PE of CBC sources have been pioneered in Ref.~\cite{Dax_2021}. While the core principle of AMPLFI is similar, there are significant implementation differences, namely

\begin{itemize}
    \item An empirical noise model that makes use of real detector data to sample noise instances for generating simulations from the likelihood (see Sec.~\ref{sec:results-robustness}).
    \item Implementation of GW signal approximants with PyTorch \cite{Bellm_2018} such that signals can be generated using GPUs on-the-fly during the training process. This implies that the training dataset is effectively infinite, and does not need to be pre-generated or saved to disk.
    \item Real-time data processing steps like filtering, power spectral density (PSD) estimation, and whitening employed on the GPU.
\end{itemize}
These algorithmic and data processing choices allow a maximally diverse, high entropy training dataset while still maintaining effective GPU utilization. In more detail, the training loop for a batch size of $N$ training examples includes

\begin{enumerate}
    \item Sampling $N$ parameter instances $\theta^{(i)}$, $i \in \{1,...,N\}$, from a prior distribution $p(\theta)$ of waveform parameters. We use the \texttt{torch.distributions} library for the priors which allows direct sampling on the GPU.
    \item Generating intrinsic signal polarizations $h_{\times}(\theta^{(i)})$ and $h_{+}(\theta^{(i)})$ in batch. This is done using signal generators implemented in \texttt{torch} and publicly available in the ml4gw library\footnote{\url{https://github.com/ML4GW/ml4gw}}.
    \item Sampling right ascension, declination and polarization angle uniformly and projecting waveforms onto the interferometers, producing the observed signal $h_k(\theta^{(i)})$ in the $k^{th}$ interferometer. 
    \item Sampling $N$ noise instances $n_k^{(i)}$ from data stored on disk, independently in time for each interferometer $k$. This includes data used for estimating the PSD (see Sec.~\ref{sec:results-robustness} for details).
    \item Injecting the waveforms into detector noise i.e., $d_k^{(i)} = h_k(\theta^{(i)}) + n_k^{(i)}$. The injection is performed in the time domain into 3 seconds of noise. The coalescence time is randomly placed with uniform probability between 0.4 and 0.6 seconds from the right edge of the window.
    \item Creating whitened data, $\tilde{d}^{(i)}_k$, using a PSD estimate $S_k^{(i)}$ local to each training sample.
    \item Jointly mapping the whitened data and PSD estimates for all interferometers, $\{\tilde{d}^{(i)}_k, S_k^{(i)}\}$, to a lower dimensional data summary $\gamma^{(i)} = \Gamma_\psi(\{\tilde{d}^{(i)}_k, S_k^{(i)}\})$ using a neural network $\Gamma$ with weights $\psi$ that are optimized during training.
    \item Providing the $N$ combinations of parameters and data summaries, $\{\theta^{(i)}, \gamma^{(i)}\}$ to the neural network, with the training objective of minimizing the negative log-probability with respect to weights $\psi$ and $\phi$.
    \begin{equation}
        \underset{\phi,\psi }{\text{min}}  \frac{1}{N}\sum_{i=1}^N -\log q_\phi\left(\theta^{(i)} \vert \gamma^{(i)}\right),
    \end{equation}
\end{enumerate}

In this work, strain data is sampled at 2048 Hz.

\subsection{Improvements} \label{sec:improvements}
\subsubsection{Training Data}
In Ref.~\cite{Chatterjee:2024pbj}, AMPLFI was trained and evaluated using a fraction of a day of strain data. As such, accounting for variability in detector noise was not critical to performance. So, a single, global estimate of the PSD was used to whiten the data for each training batch. When deployed to a production environment, any deviations in noise properties could lead to model under performance. Practically, using a global PSD estimate encoded the assumption of noise-stationarity into the model. 

In this work, the quantity of strain data used for training is increased to $\sim$ 2 months. A global PSD estimate is no longer sufficient at modeling the varying noise profiles of the detectors across this long of a period. Instead, we now utilize data local to each training sample to estimate the PSD. For each training noise instance, 64 seconds of strain data immediately preceding the sample is used to estimate the PSD using Welch's method with median averaging. Increasing the quantity of strain data and using local PSD estimates allows a richer sampling of detector noise states, leading to improved generalization and model longevity. This amount of data cannot fit into memory at once. We take advantage of an efficient out-of-memory data loader implemented in ml4gw so that data loading does not bottleneck the training process. This is same data loader implementation is used to train the Aframe detection algorithm \cite{aframe-methods}.

Lastly, in Ref.~\cite{Chatterjee:2024pbj}, AMPLFI was trained with a GPU-accelerated IMRPhenomD approximant \cite{phenomD}. Since then, we have implemented the IMRPhenomPv2 waveform approximant, allowing AMPLFI to be trained with a waveform that includes spin precession physics.

\subsubsection{Neural Network}
In Ref.~\cite{Chatterjee:2024pbj}, the AMPLFI embedding network $\Gamma_\psi$ consisted of a 1 dimensional convolutional ResNet \cite{resnet}, which processed time-domain representations of the data to produce a lower-dimensional data summary. Here, we introduce a parallel embedding network that processes a frequency-domain representation of the signal and a local PSD estimate, $S$. The data summaries produced by the time and frequency embedding networks, denoted $\gamma_t$ and $\gamma_f$, respectively, are concatenated into a final data summary, $\gamma$, which is used to condition the normalizing flow.
\begin{align}
    \gamma_t &= \Gamma_{\psi_t}^{t}(\tilde{d}) \\ \label{eq:gamma_t}
    \gamma_f &= \Gamma_{\psi_f}^{f}(FFT(\tilde{d}), S) \\ \label{eq:gamma_f}
    \gamma &= \text{Concat}(\gamma_t, \gamma_f)
\end{align}
The real and imaginary parts of the frequency domain data are processed as separate channels by the frequency domain embedding network. In equations \ref{eq:gamma_t} and \ref{eq:gamma_f}, $\psi_t$ and $\psi_f$ correspond to the trainable parameters of the time domain and frequency domain embedding networks, $\Gamma_{\psi_t}^{t}$ and $\Gamma_{\psi_f}^{f}$, respectively. The dimensions of $\gamma_t$ and $\gamma_f$ are hyperparameters that can be optimized, and control how much information from each domain is passed to the normalizing flow.

Providing redundant data in the form of the Fourier transform can be interpreted as an inductive bias. Intuitively, different data representations provide easier pathways for the neural network to learn certain parameters. As an example, the coalescence time might be easier to identify in the time domain, whereas the chirp mass, a quantity determined by a power law in frequency, might be easier in the frequency-domain. Multi-modal learning approaches have been explored in other GW data analysis contexts \cite{Bhardwaj:2023xph, Sun:2023vlq}.

In addition, we also increase the number of trainable parameters of the normalizing flow. In, Ref.~\cite{Chatterjee:2024pbj}, affine flow transforms were used \cite{realnvp}. Here, we use the more expressive spline-based transforms \cite{spline}. We reduce the number of auto-regressive transforms, and compensate by increasing the hidden dimensions and number of hidden layers of the hyper-networks used to learn the parameters of the spline transformations. Normalizing flows are implemented with the zuko library \cite{rozet2022zuko}. 

For the studies that follow, we train two AMPLFI models for Hanford-Livingston (HL) and Hanford-Livingston-Virgo (HLV) detector configurations. Both HL and HLV neural networks utilize the same fundamental architecture described above. However, the HLV network utilizes a larger normalizing flow component. See Table~\ref{tab:hps} for hyperparameters used in this work. For each detector configuration, we use coincident science mode segments from the beginning of the second half of the third observing run (O3b) until the start of the testing dataset (see Sec.~\ref{sec:testing-data}) for training. We train the models using the low-latency calibrated strain data. The embedding and normalizing flow for the HLV (HL) model total 70 (30) million parameters, a factor of $\sim 10$ increase from Ref.~\cite{Chatterjee:2024pbj}. Training this larger model takes $\sim 10$ days using 2 NVIDIA A100 GPUs.

\begin{table}[h]
    \begin{tabular}{lcc} 
    \hline
    Hyperparameter & HL & HLV \\
    \hline\hline
    \# Transforms & 15 & 20 \\
    \hline
    \# Hyper-network layers & 3 & 3 \\
    \hline
    \# Hidden units & 1024 & 1024 \\
    \hline
    size of $\gamma_f$  & 32 & 48 \\
    \hline
    size of $\gamma_t$ & 12 & 20 \\
    \hline
    \end{tabular}
    \caption{Neural network hyperparameters for the HL and HLV models used in this work.}
    \label{tab:hps}
\end{table}

\section{Testing Dataset} \label{sec:testing-data}
In preparation for O4, Ref.~\cite{Chaudhary:2023vec} performed a study which added simulated CBC signals (known colloquially as \emph{injections}) into a real-time data replay. The injections were added across a one month period during O3b, between 2020-01-05T:23:59:42 and 2020-02-14T23:59:42. Search pipelines analyzed the data replay using their online configurations. GWCelery\footnote{\url{https://git.ligo.org/emfollow/gwcelery}}\footnote{\url{https://rtd.igwn.org/projects/gwcelery/en/latest/}}processed pipeline triggers, mimicking the end-to-end real-time alert infrastructure. This study provided matched-filter pipeline SNR time series for injections across a broad parameter space. 

During the real-time analysis, a given injection may be detected by multiple different matched-filter pipelines, creating several associated SNR time series to use for localization. For comparison, we choose the localization corresponding to the preferred event in GraceDB. This corresponds to the matched-filter event with the highest SNR. In some instances, the preferred event is associated with the un-modeled cWB pipeline which does not produce sky localizations with BAYESTAR. For these instances, we select the matched-filter event with the highest SNR. 

\begin{figure}
    \centering
    \includegraphics[width=0.95\columnwidth]{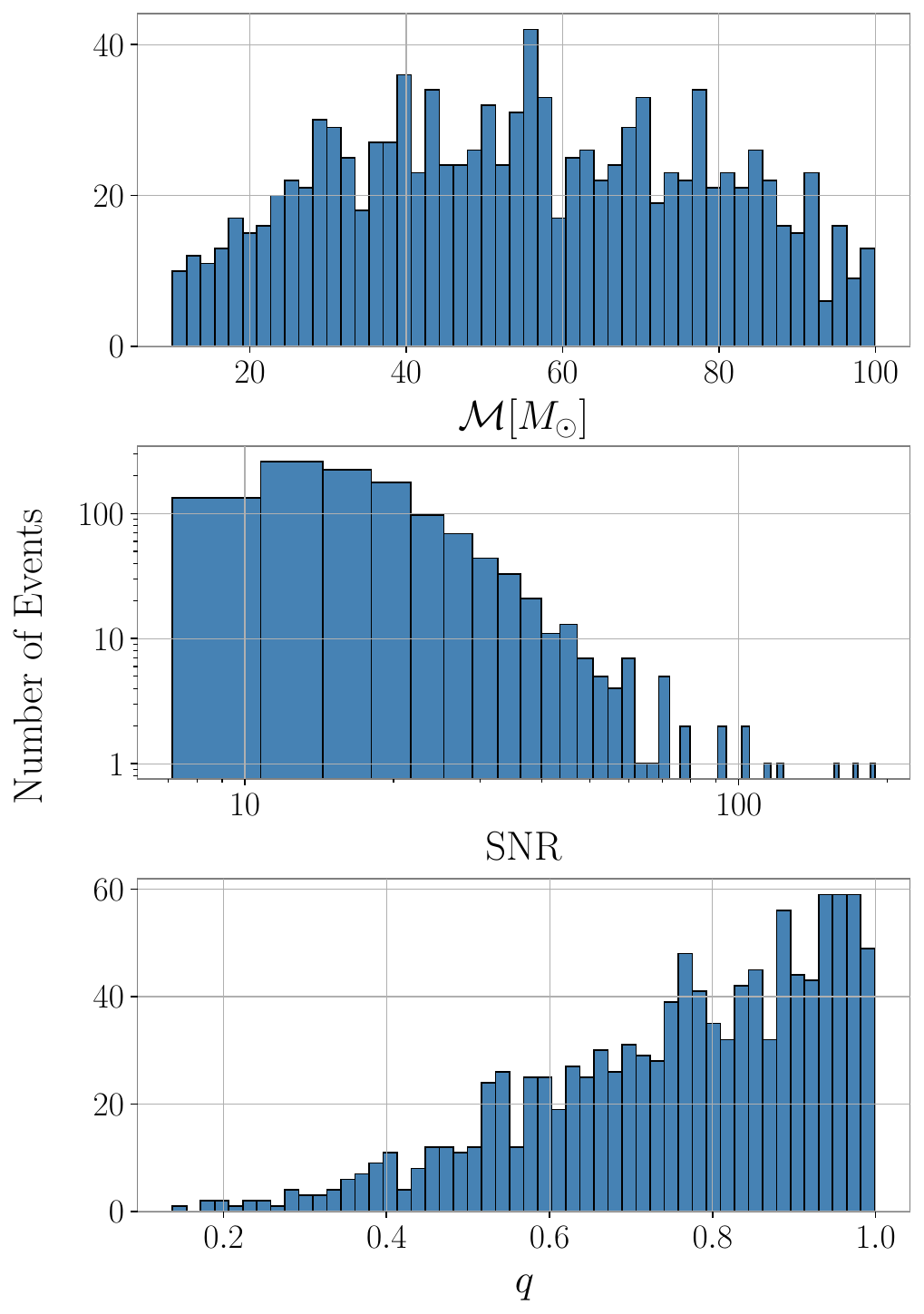}
    \caption{Histograms of chirp mass (top), SNR (middle) and mass ratio (bottom) for the 1233 HL injections used in this study to compare with BAYESTAR. The HLV injections are a strict subset of the HL injections.}
    \label{fig:mdc_dist}
\end{figure}

Using these SNR time series, we create a dataset of BAYESTAR localizations. We first filter the injections to those consistent with AMPLFI's training prior (see Table~\ref{tab:priors}). The injections are binary black hole mergers with chirp masses between 10 and 100 $M_{\odot}$. We ensure the injections were performed using science quality data from each interferometer. Using the SNR time series, we create Hanford-Livingston (HL), and if available, Hanford-Livingston-Virgo (HLV) BAYESTAR localizations. BAYESTAR is configured to use the same distance prior used to train AMPLFI. After these cuts, this leaves 1233 HL and 903 HLV injections.  Figure~\ref{fig:mdc_dist} visualizes the distribution of signal parameters. The dataset consists of a broad parameter space of signals, nearly uniform in chirp mass between 10 and 100 $M_\odot$ and covering a wide range of SNRs.

\begin{table}
\centering
\renewcommand{\arraystretch}{1.1}
\begin{tabular}{p{3cm} p{5cm}}
\hline
\textbf{Parameter} & \textbf{Prior} \\
\hline
$a_{1,2}$ \hfill (Spin magnitude) & Uniform(0, 0.999) \\
$\phi_{12}$ \hfill (Spin azimuthal angle) & Uniform(0, $2\pi$) \\
$\phi_{jl}$ \hfill (Spin phase angle) & Uniform(0, $2\pi$) \\
$\text{tilt}_{1,2}$ \hfill (Spin tilt angle) & Uniform(0, $\pi$) \\
\hline
\multicolumn{2}{l}{Inference} \\
\hline
$\mathcal{M}_c$ \hfill (Chirp mass) & Uniform(10, 100) $M_{\odot}$ \\
$q$ \hfill (Mass ratio) & Uniform(0.125, 1) \\
$d_L$ \hfill (Luminosity distance) & Uniform (100, 3100) Mpc \\
$\theta_{jn}$ \hfill (Inclination) & Sine(0, $\pi$) \\
$\alpha$ \hfill (Right ascension) & Uniform(0, $2\pi$) \\
$\delta$ \hfill (Declination) & Cosine($-\pi/2$, $\pi/2$) \\
$\phi_c$ \hfill (Coalescence phase) & Uniform(0, $2\pi$) \\
$\psi$ \hfill (Polarization angle) & Uniform(0, $\pi$) \\
\hline
\hline
\end{tabular}
\caption{Prior distributions for source parameters used to generate waveforms during training. Parameters which AMPLFI is trained to infer are specified. We use a spin parameterization consistent with Bilby. Note that the distance prior is uniform in distance. So, cosmological effects are not included, but can be incorporated in post-processing through importance sampling.}
\label{tab:priors}
\end{table}

\section{Sky Map Production} \label{sec:sky-localization}
GW sky maps are probability density distributions on the sky constructed using the HEALPix\footnote{\url{http://healpix.sourceforge.net}} format \cite{Zonca2019}. For CBC events with distance calibrated waveform models, three-dimensional volume information can also be estimated. BAYESTAR estimates the parameters of a conditional distance distribution ansatz for each sky pixel which is distributed as part of the sky map in low-latency GW discovery alerts \cite{Si2016}. 

For stochastic sampling algorithms which draw samples from the posterior, a method for estimating each sky pixels probability density and distance ansatz parameters from the samples is required. Here, we compare two techniques for estimating the densities and distance ansatz parameters: a kernel density estimate (KDE) using the \texttt{ligo-skymap-from-samples} tool, and an adaptive histogram estimator which adaptively grids the sky, creating higher resolutions in regions with higher numbers of posterior samples. Both of these tools are available via the \texttt{ligo.skymap}\footnote{\url{https://git.ligo.org/lscsoft/ligo.skymap}} library. We show that the searched area performance for the two methods is equivalent. For searched volume, the performance is comparable at mid to high SNRs. However, the KDE begins to outperform the adaptive histogram estimator at low SNRs ($\lesssim 10$).

The adaptive histogram density is estimated as follows

\begin{enumerate}
    \item Draw approximately $10^4$ samples from the normalizing flow. This step takes $\sim 1-2$ seconds on an NVIDIA A30 GPU.
    \item Bin the samples using an adaptive HEALPix grid. 
    Resample the grid to a flat resolution with NSIDE=64.
    This corresponds to pixel area of $\sim 1\;\mathrm{deg.}^2$ and
    provides sufficient resolution for all-sky survey instruments which have fields
    of view $1-50\;\mathrm{deg}^2$ \cite{andreoni2024rubin2024envisioningvera, Bellm_2018}.
    \item Using the posterior samples in each sky pixel, calculate the first two distance moments,
    and solve for the distance distribution ansatz parameters as described in Ref.~\cite{Singer_2016}.
    \item De-rasterize the sky map into multi-order scheme \cite{Fernique}. This reduces the data size of the sky map with no information loss, allowing for low-latency distribution via alert brokers (e.g. GCN and SCiMMA\footnote{\url{https://scimma.org/}}).
\end{enumerate}
We calculate the conditional distance ansatz parameters for pixels with $\geq 5$ samples. For pixels which have a smaller number or no samples, we use placeholder values as done in Ref.~\cite{Singer_2016}. The adaptive histogram estimator including distance estimation can be run in less than a second.

The KDE is constructed by modeling p($\alpha, \delta, d_L$) as the product of the 2D marginal distribution p($\alpha, \delta$) and the conditional distance distribution p($d \vert \alpha, \delta$). For both, posterior samples are clustered using a k-means algorithm into $k$ clusters. A brute-force search over $k$ is performed in order to maximize the Bayesian information criterion. For each cluster, a separate KDE is constructed, and the total distribution is modeled by summing the cluster KDEs with appropriate weights. The conditional distance distribution p($d \vert \alpha, \delta$) is evaluated as $p(\alpha, \delta, d_L) / p(\alpha, \delta)$ by analytically marginalizing the 3D KDEs. See Sec.~5 in \cite{Singer_2016} for more details. Due to the brute-force search over $k$, the runtime is $\sim 11$ seconds parallelized on 64 CPUs, which is not suitable for extremely rapid alerts.

\begin{figure} 
    \includegraphics[width=0.95\columnwidth]{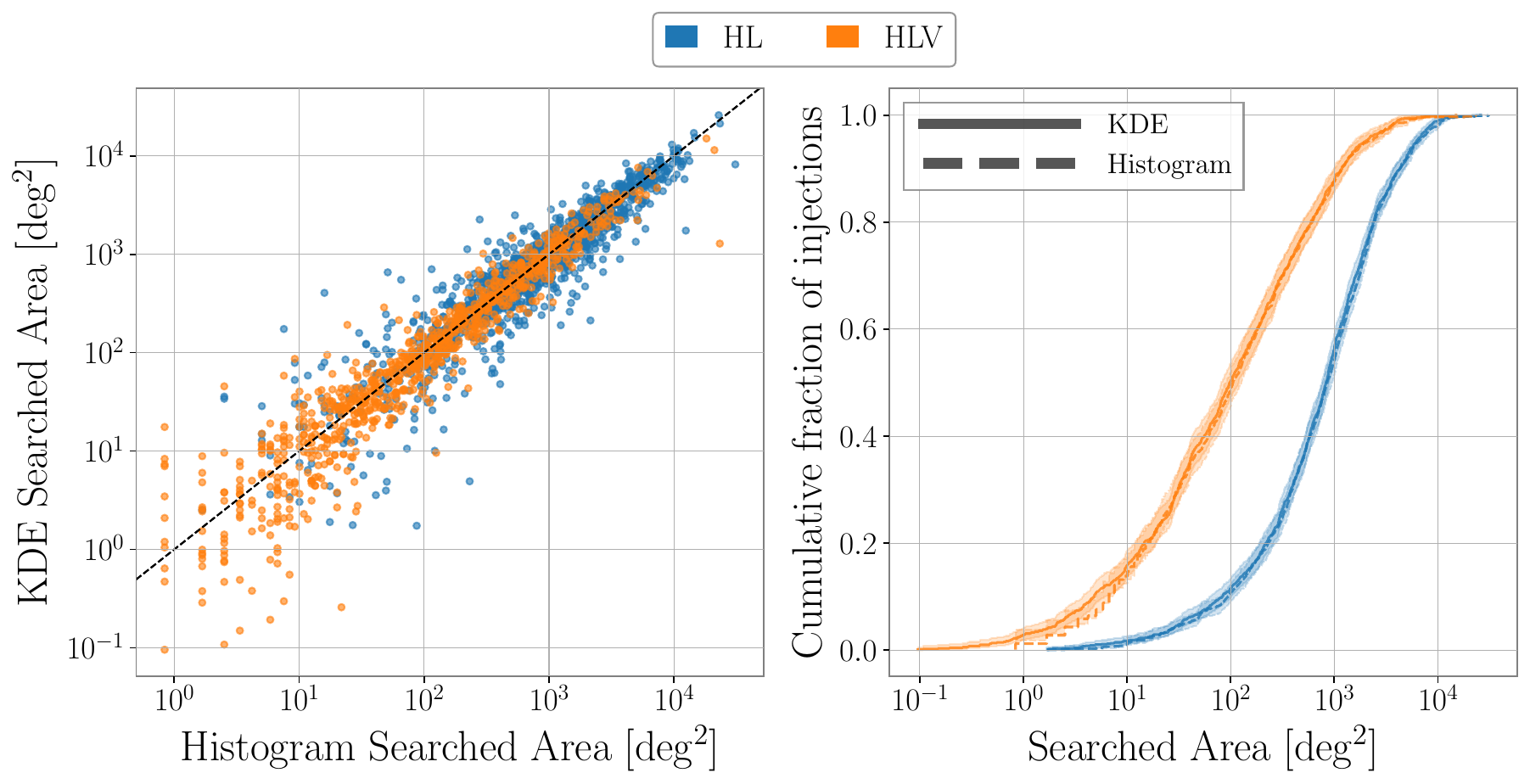}
    \includegraphics[width=0.95\columnwidth]{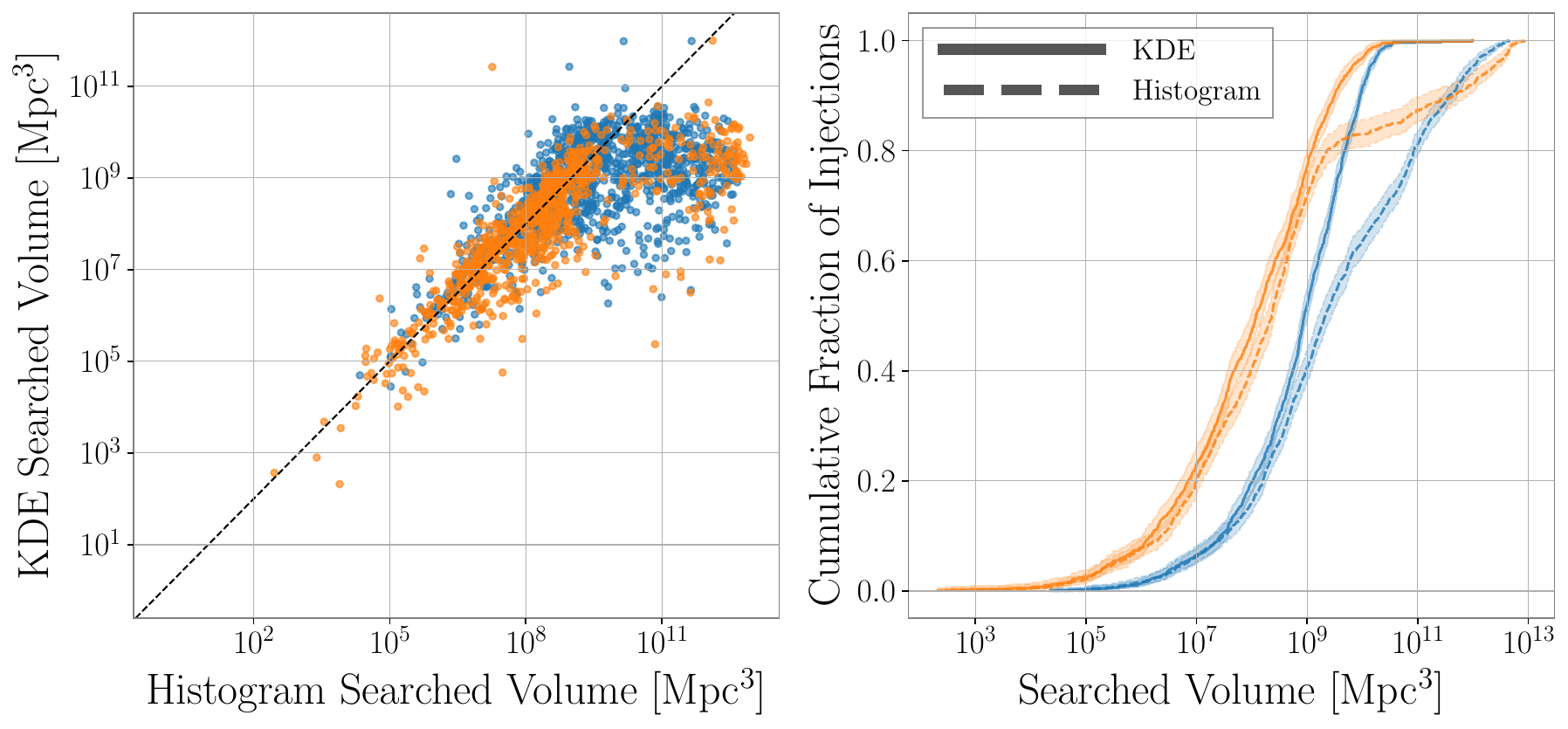}
    \caption{Top Left: Scatter plot of searched areas comparing the histogram and KDE estimators for the HL (blue) and HLV (orange) injections. Top Right: Cumulative histogram of searched areas for the same events. Error bars correspond to 1$\sigma$ confidence intervals calculated by bootstrapping are shown. Searched areas from the two density estimation methods are in statistical agreement. Bottom: Same as top, but for searched volumes. Searched volumes for the two methods broadly agree up until $\sim10^{9}$ $\mathrm{Mpc}^3$ for HLV and $\sim 10^{8}$ $\mathrm{Mpc}^3$ for HL network , where the KDE method begins to outperform the histogram estimator.}
    \label{fig:kde-vs-hist}
\end{figure}

Using the dataset of injections described in Sec.~\ref{sec:testing-data}, we compare the accuracy of the AMPLFI sky map produced using the histogram and KDE estimators. For each injection, we calculate searched area and searched volume metrics for both estimators. Searched area (volume) is measured by sorting sky area (volume) pixels by descending probability, and accumulating area (volume) until the pixel containing the true location of the simulated signal is reached. For the adaptive histogram estimator, we draw 20,000 samples from the model. The top row of Figure~\ref{fig:kde-vs-hist} shows that searched areas from the two methods are in statistical agreement. However, the histogram estimator has a minimum searched area of $\sim 1$ $\mathrm{deg^2}$ due to the pixel resolution of NSIDE=64, and is therefore unable to achieve searched areas smaller than this. The bottom row compares the searched volumes between the estimators. The methods agree for searched volumes $\lesssim 10^9$ $\mathrm{Mpc}^3$. For larger searched volumes, the KDE method performs better. This behavior is due to the large number of samples required for the histogram estimator to properly converge for low SNR events, where the sky localization probability is spread across a larger number of pixels. In practice, these low SNR events are not as likely to be pursued for EM followup. For example, the Rubin target of opportunity observing strategy plans to pursue EM followup for events with sky localizations less than $\sim 100\;\text{deg}^2$ \citep{andreoni2024rubin2024envisioningvera}. In this regime, the histogram methods searched volume is comparable with the KDE.

\begin{figure*}
\includegraphics[scale=0.5]{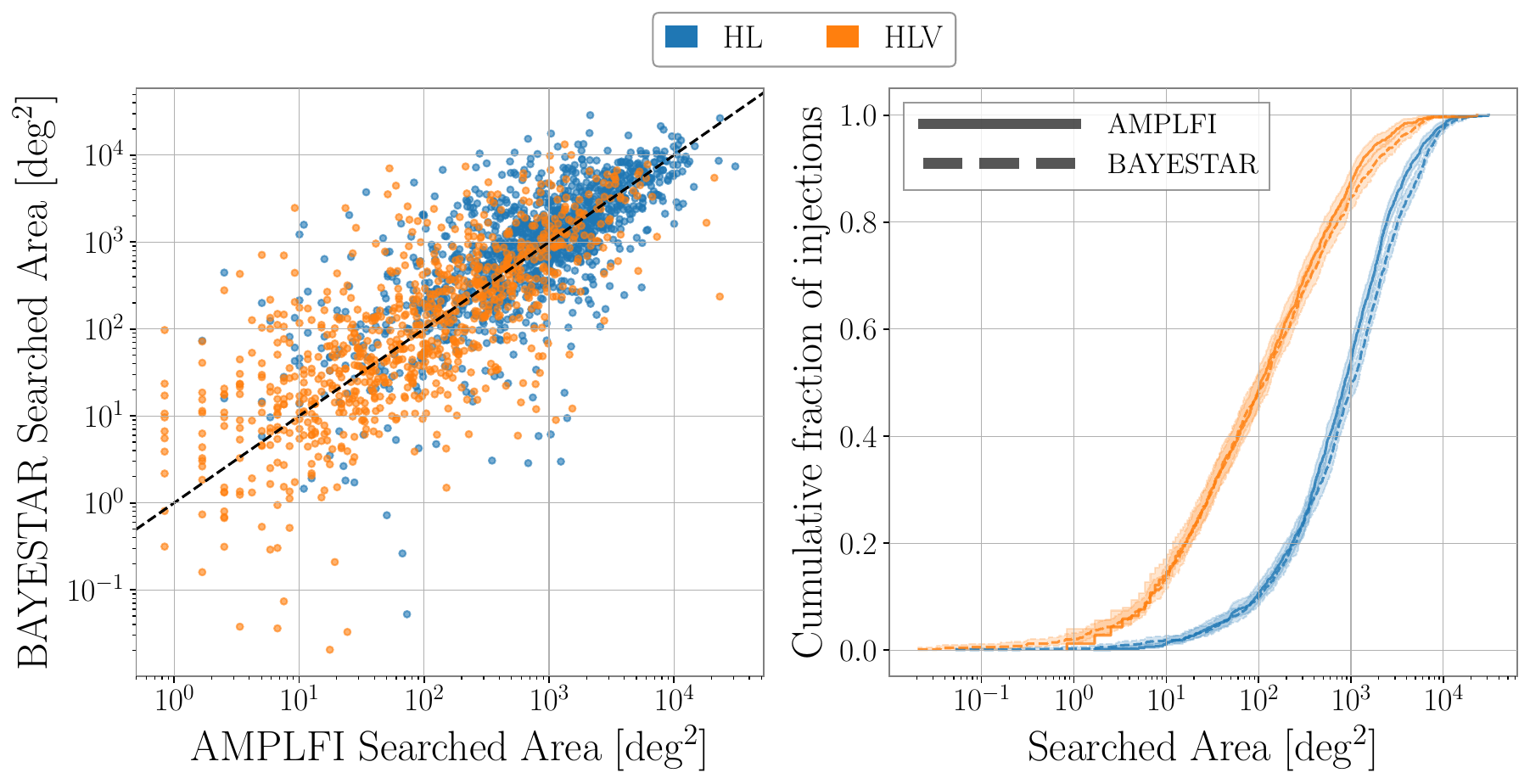}
\includegraphics[scale=0.5]{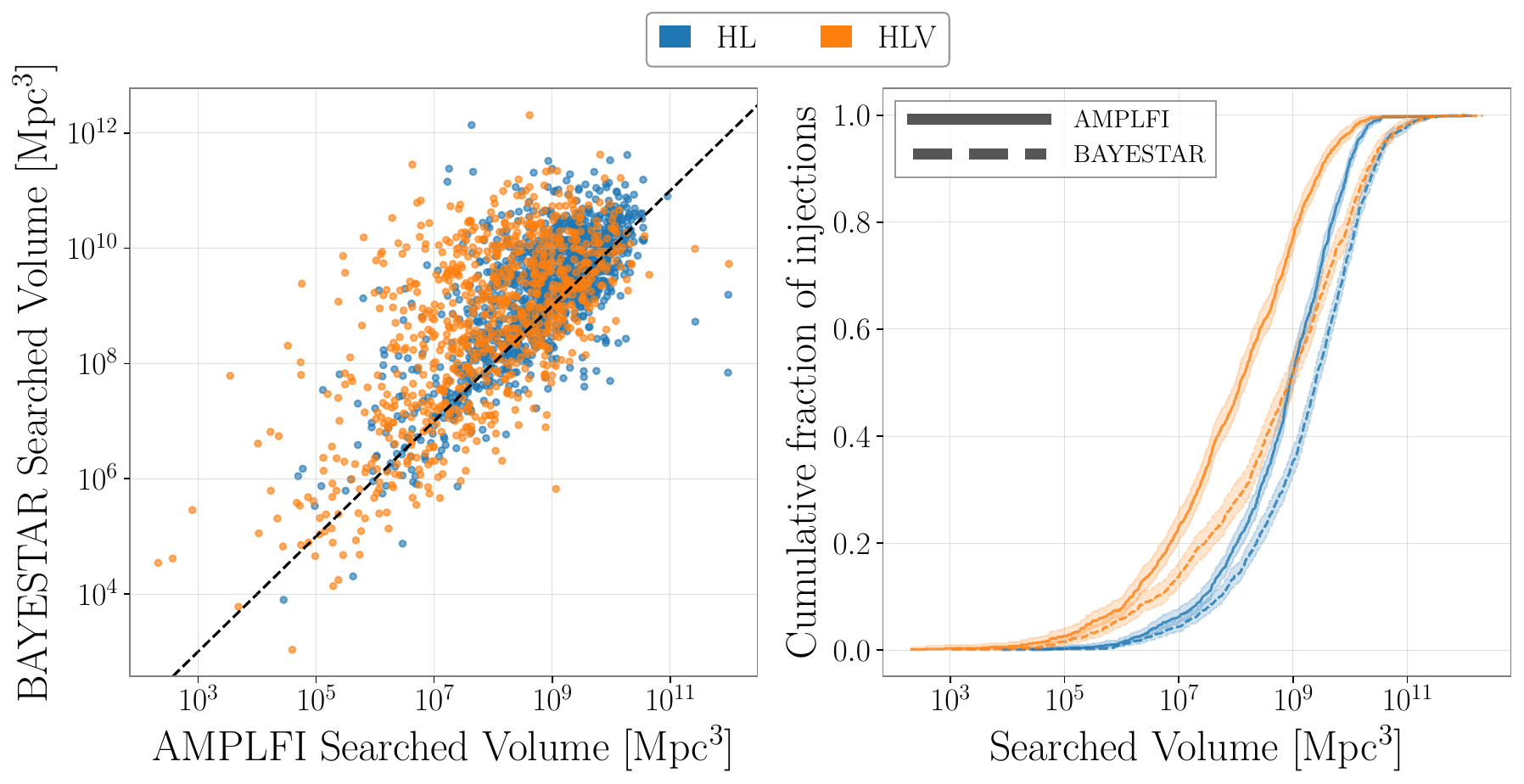}
\caption{Top Left: Scatter plot comparing searched areas of AMPLFI and BAYESTAR. HLV injections are show in orange, and HL injections are shown in blue. Top Right: Cumulative histogram of searched areas for the same events. Error bars correspond to 1$\sigma$ confidence intervals calculated by bootstrapping are shown. The cumulative distribution of searched areas from the two methods are statistically equivalent. Bottom: Same as top, but for searched volumes. AMPLFI provides smaller searched volumes than BAYESTAR} \label{fig:amplfi-vs-bayestar}
\end{figure*} 

\section{Performance Studies} \label{sec:results}
 
\subsection{Sky Maps} \label{sec:results-skyloc}

\begin{figure}
    \centering
    \includegraphics[width=\columnwidth]{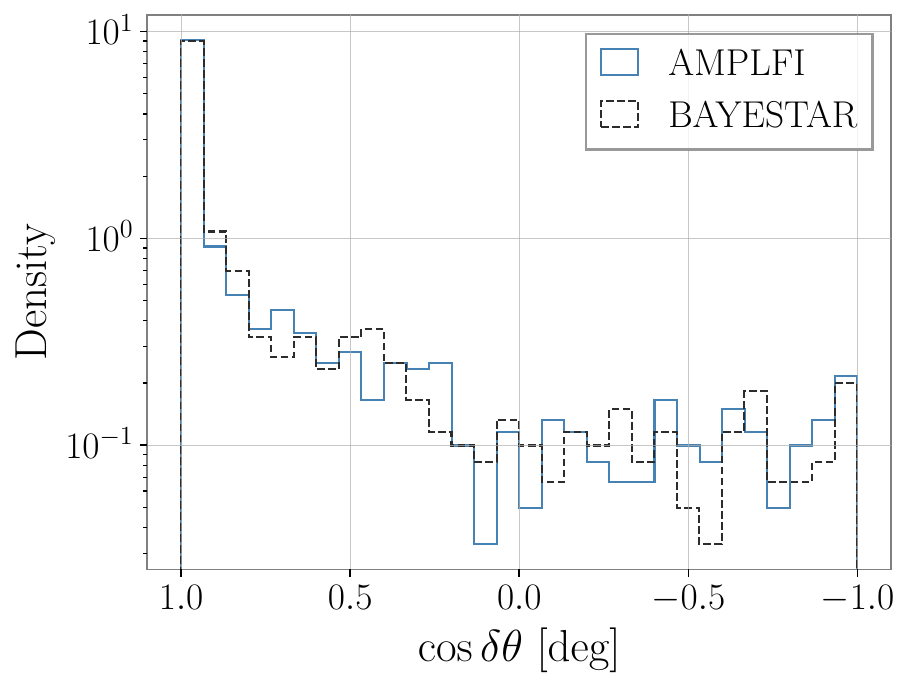}
    \caption{Normalized histogram of cos $\delta \theta$ for the HLV injection set, where $\delta \theta$ is the angular offset between the true pixel of the injection and the maximum a posterior pixel}
    \label{fig:dtheta}
\end{figure}

\subsubsection{Accuracy}
We compare AMPLFI's sky localization performance with BAYESTAR using the dataset set described in Sec.~\ref{sec:testing-data}. In Figure~\ref{fig:amplfi-vs-bayestar} we compare searched areas (top) and searched volumes (bottom) between AMPLFI and BAYESTAR sky maps for both HLV and HL datasets. Bootstrapped confidence intervals show the searched area cumulative histograms agree within 2$\sigma$. The searched volume cumulative histogram shows that AMPLFI provides smaller searched volumes than BAYESTAR. One drawback of searched area and volume as performance metrics is they do not capture sky map multi-modality (see Ref.~\cite{Essick:2014wwa} for a discussion). In Figure~\ref{fig:dtheta} we compare the angular offset between the sky maps maximum a posterior pixel and the true pixel of the injection. We plot $\cos \delta \theta$ which highlights groupings at the true and antipodal location. We see agreement between the BAYESTAR and AMPLFI distributions.

\subsubsection{Consistency} \label{sec:results-consistency}
Searched area and volume measure accuracy, but not calibration (i.e. probabilistic consistency). Sky map calibration can be evaluated by performing probability-probability (P-P) tests on a dataset of simulated signals. For well calibrated sky maps, reported confidence intervals should correspond to the probability of finding the true location of the GW source in that area or volume region. This probability can be estimated empirically by analyzing injections. 

For BAYESTAR, this calibration is achieved by rescaling the SNR timeseries by a correction factor which tuned to injections \cite{bayestar}. Notably, this correction factor is dependent on algorithmic details of the matched-filter pipeline that provides the SNR timeseries. The default value of 0.83 is tuned to injections analyzed by the GstLAL pipeline. Using this default correction factor, BAYESTAR sky maps produced from the SNR timeseries provided by the PyCBC live analysis have been shown to be biased (see Figure 5 in Ref.~\cite{Nitz_2018}). The source of this correction factor was further studied in Ref.~\cite{Duverne_2024} using PyCBC. The correction factor was shown to depend on variables like the construction of the matched-filtering template bank, and detector configuration. The variability of this correction factor upon pipeline specifics is undesirable, since it requires tuning to algorithmic choices.

Figure~\ref{fig:self-consistency} shows self-consistency tests for the HLV AMPLFI model using injections with SNR $\geq 12$. Consistency tests for area and volume are shown to be unbiased. A small bias in distance is present. Since AMPLFI is trained directly on strain data it does not require additional pipeline dependent tuning factors to achieve self consistency.

\subsection{Intrinsic Parameter Recovery}
\begin{figure}
    \centering
    \includegraphics[scale=0.35]{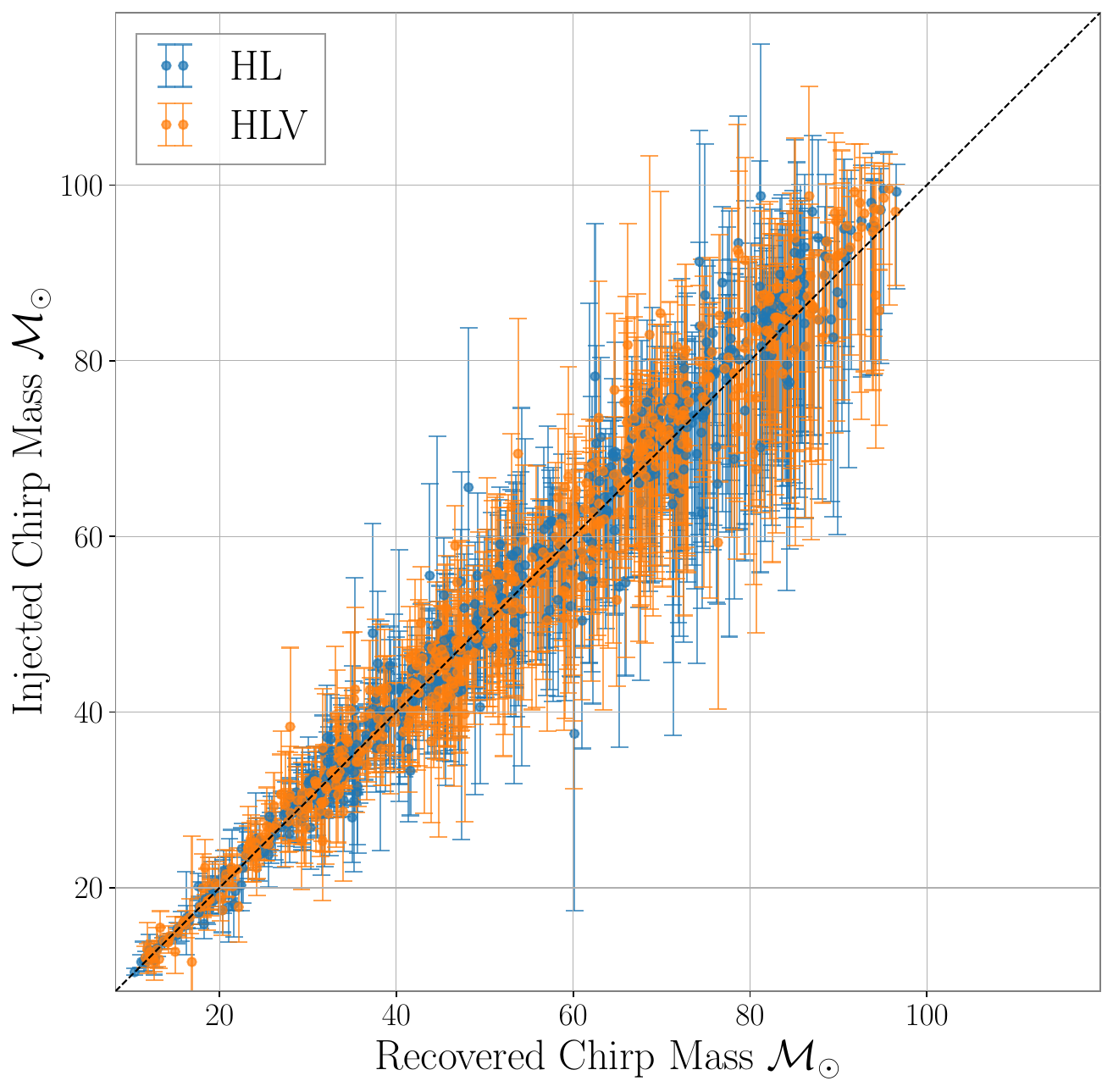}
    \caption{Scatter plot comparing true chirp mass and AMPLFI's median chirp mass for simulated signals from the injection dataset described in Sec.~\ref{sec:testing-data}. Error bars corresponding to the $5^{\mathrm{th}}$ and $95^{\mathrm{th}}$ percentiles of AMPLFI's posterior samples are shown. Orange and blue points correspond to the HL and HLV networks respectively.} 
    
    \label{fig:chirp-mass-recovery}
\end{figure}

\begin{figure*}
    \centering
    \includegraphics[width=\textwidth]{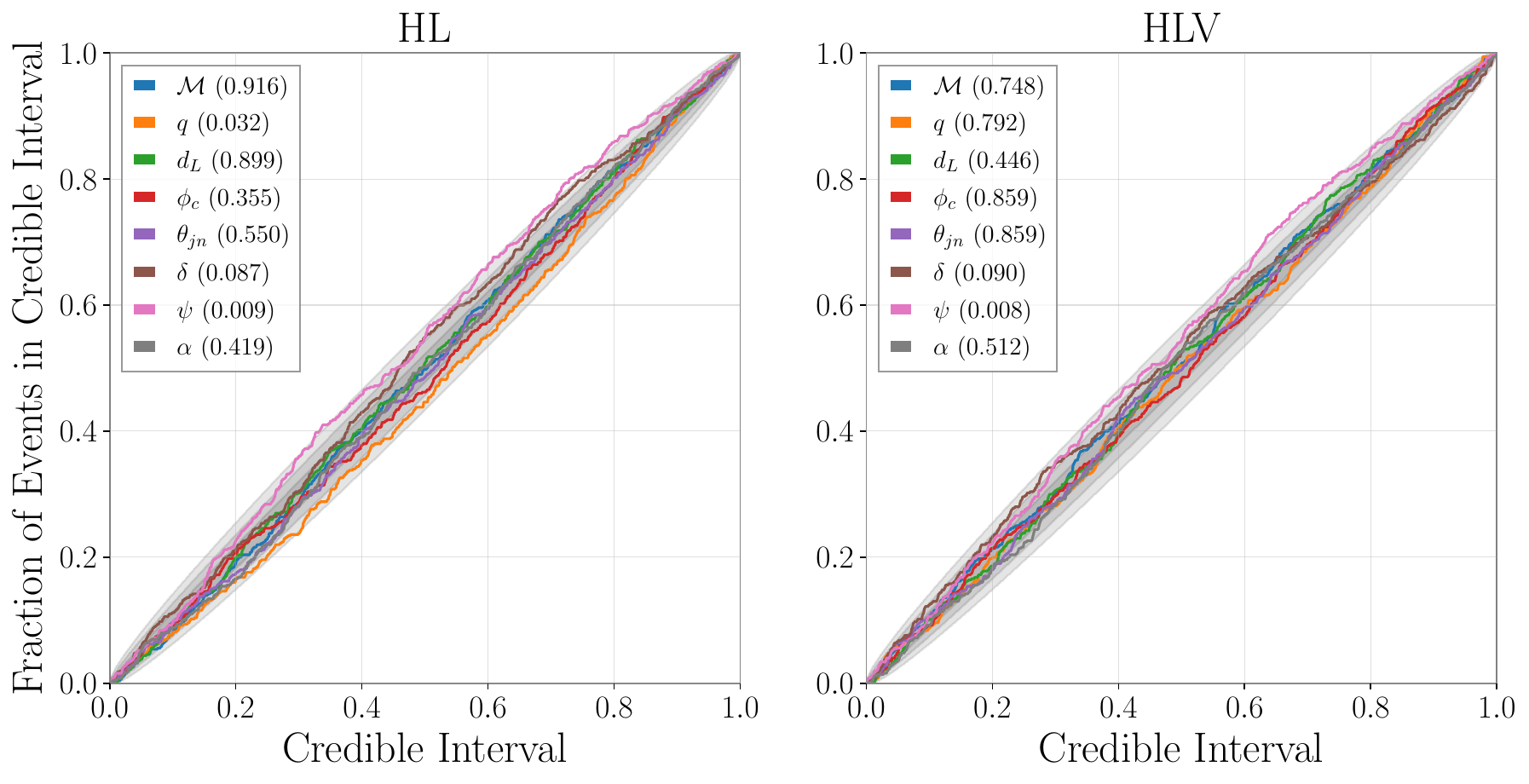}
    \caption{P-P tests for AMPLFI's 1 dimensional marginal posterior distributions. 500 injections from AMPLFI's training prior were analyzed. The p-value for each parameter is provided in the legend. Shaded bands correspond to 1, 2 and 3 $\sigma$ confidence intervals.}
    \label{fig:pp-plot}
\end{figure*}

\begin{figure*}
    \centering
    \includegraphics[width=\textwidth]{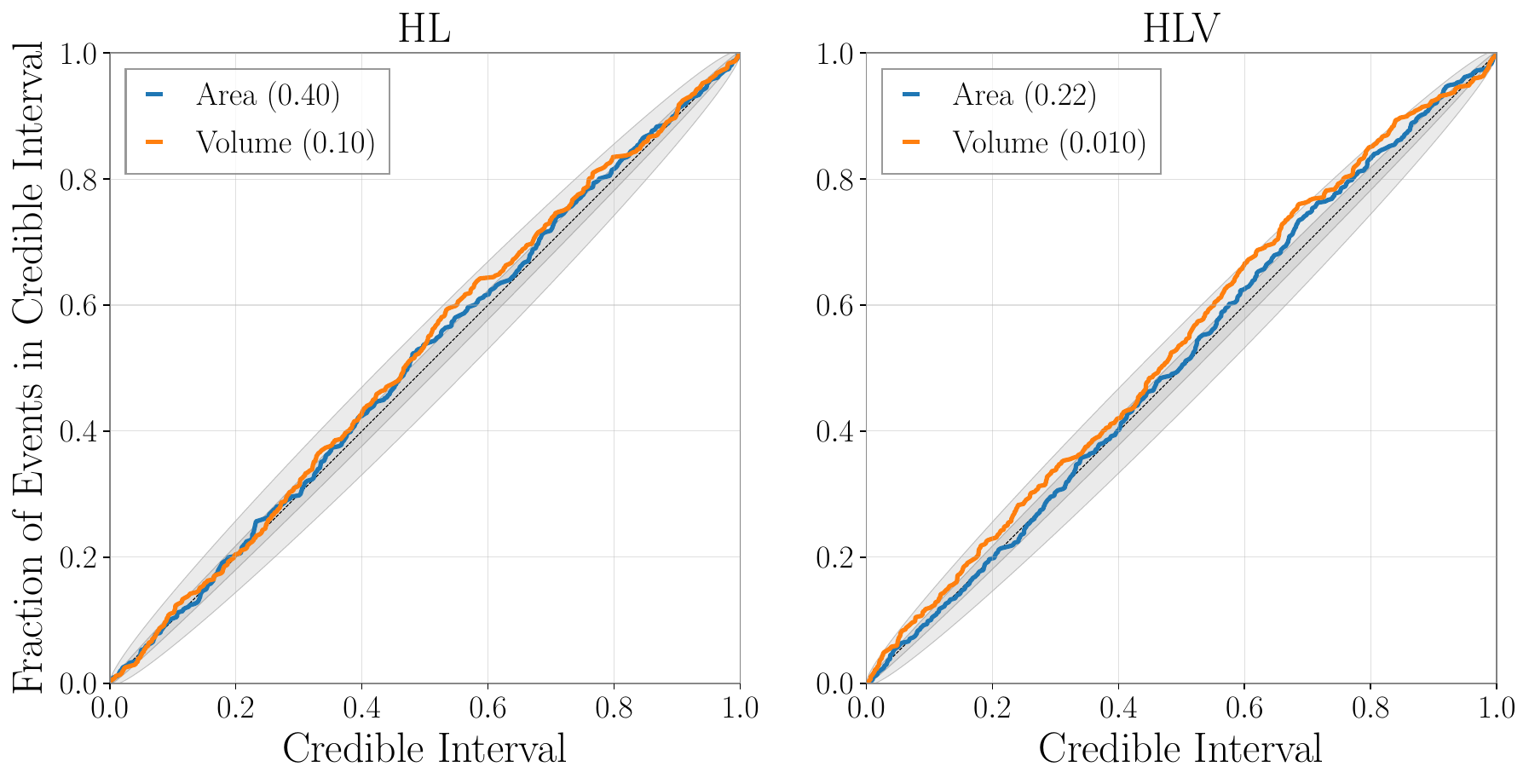}
    \caption{P-P tests for AMPLFI's joint 2D (area) and 3D (volume) posterior distributions measuring sky map self consistency. The p-value for each consistency check is provided in the legend. Error bars corresponding to 1 and 3 $\sigma$ confidence intervals are plotted as shaded bands.}
    \label{fig:self-consistency}
\end{figure*}

In low-latency, matched-filter pipelines provide only point estimates of source parameters. These point estimates can suffer from systematic biases due to algorithmic choices such as template bank construction \cite{Ewing:2023qqe}. These biases have been shown to affect the performance of downstream source property classifiers which are used to publicly distribute probabilities of the source containing a neutron star, remnant, or mass gap component \cite{Chatterjee:2019avs}. Because AMPLFI provides a full posterior, estimates of uncertainty can be used in low latency. These estimates can be used to restrict the prior space for downstream PE tasks which aim to explore the full signal parameter space including spins. Restricting the prior space leads to faster PE using fewer computational resources.
In Fig.~\ref{fig:chirp-mass-recovery}, we illustrate AMPLFI's ability to recover the chirp mass of the source. We see the expected increase in uncertainty as the injected chirp mass increases. In addition, we perform a P-P test using 500 injections drawn from AMPLFI's training prior. Fig.~\ref{fig:pp-plot} shows a P-P test demonstrating AMPLFI's unbiased recovery of parameters.

\subsection{Model Robustness} \label{sec:results-robustness}
GW interferometers have non-stationary noise distributions \cite{Davis_2021, LIGO:2024kkz}. Changing noise distributions pose problems to machine learning algorithms, which can suffer from overfitting to the noise statistics of the data period used for training. For production grade deployment, it is critical to validate the robustness of model performance beyond the initial training period, and the length of time over which satisfactory performance is maintained. 

Different strategies can be employed to construct a robust noise model. For example, in Ref.~\cite{Wildberger_2023}, data augmentations applied to a fiducial PSD are utilized to construct a noise model that captures the variation in noise properties across an observing run. In that work, the augmented PSDs are used to generate synthetic Gaussian noise for training. As described in Sec~\ref{sec:review}, AMPLFI utilizes real detector strain data for training, and so our noise model is constructed empirically. Specifically, our training data consists of T disjoint, coincident, science quality segments $\{[a_i, b_i]\}_{i=1}^{T}$.  Let $l$ denote the length of each training sample provided to the neural network, and $s$ denote the length of data used for PSD estimation. The algorithm for sampling strain data instances during training is as follows. For each detector $k$,
\begin{enumerate}
    \item Sample segment index $i_k$, with probability proportional to the segments duration:
    \begin{align*}
       p(i_k) = \frac{b_{i_k} - a_{i_k}}{\sum_{j=1}^T{(b_j - a_j)}}
    \end{align*}
    
    \item Sample time $\tau_k \sim \mathrm{Unif}([a_{i_k} + s,b_{i_k + 1} - l])$
    \item Select strain data for detector $k$ from $\tau_k - s$ to $\tau_k + l$ 
\end{enumerate}
This procedure is repeated $N$ times, where $N$ is the size of the training batch. 

The idea behind this sampling procedure is that noise between different detector sites is highly uncorrelated. So, strain data sampled independently in time for each detector is a plausible, unique noise realization. This idea, well-known in the GW literature as time-slides, is commonly used to create realistic noise realizations for estimating the background distributions of search algorithms. During training, the detector strain data is stored on disk and the above sampling procedure is done on-the-fly.

\begin{figure*}
    \centering
    \includegraphics[scale=0.5]{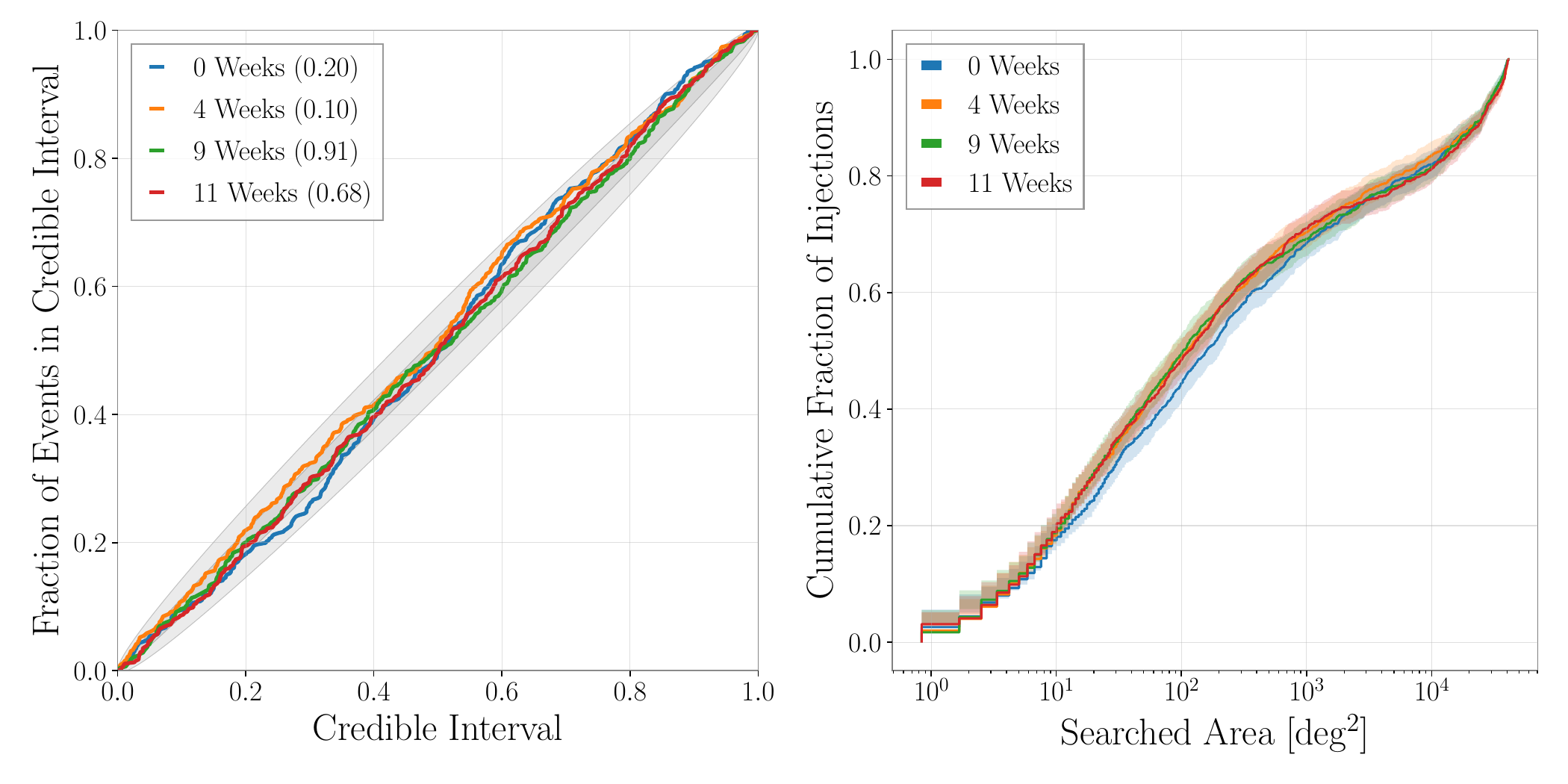}
    \caption{Left: 2D (area) localization P-P tests for 4 datasets of 1000 injections drawn from AMPLFI's training prior created at various epochs across O3b. P-values shown in the legend indicate that sky maps produced by AMPLFI remain well calibrated across an 11 week period. Right: Cumulative histogram of searched area for the same datasets. Bootstrapped 95\% confidence intervals are plotted as shaded bands. The accuracy of AMPLFI sky maps does not degrade over time.}
    \label{fig:robustness}
\end{figure*}

To validate AMPLFI's robustness over time, we created 4 datasets of 1000 injections sampled from AMPLFI's training prior and added them into real detector noise. For each dataset, we utilized strain data at different epochs after the initial model training period. These periods spanned from immediately after, to 11 weeks after the training period. Each period consisted of $\sim 1$ day of live time. We analyze each of these datasets using the HLV AMPLFI network. In Fig~\ref{fig:robustness}, we show sky localization consistency and searched area performance. We see that AMPLFI's sky localizations remain well calibrated and accurate 11 weeks beyond AMPLFI's initial training period. This demonstrates that a single AMPLFI model maintains utility well beyond its initial training period. 

\subsection{Analysis of GWTC-3 Events} \label{sec:results-gwtc3}
Several GW candidates with probability of astrophysical origin greater than $0.5$ were identified in the third Gravitational-Wave Transient Catalog (GWTC-3) \cite{Abbott_2023}. We analyze those candidates with posterior support within AMPLFI's training prior (Table \ref{tab:priors}), and which occurred during or after the period of data used to train AMPLFI. In addition, we do not analyze candidates which required glitch mitigation. After these constraints, this leaves 5 candidates analyzed with the HLV network (\Cref{fig:GW191215_223052-hlv,fig:GW200129_065458-hlv,fig:GW200208_130117-hlv,fig:GW200224_222234-hlv,fig:GW200311_115853-hlv}), and 3 candidates analyzed with the HL network (\Cref{fig:GW191204_110529-hl,fig:GW200225_060421-hl,fig:GW200306_093714-hl}). We compare posteriors and sky maps with those produced by the GWTC-3 analysis using samples from the IMRPhenomXPHM waveform. Where available, we also compare the low-latency sky map produced by BAYESTAR. Analyzing each HLV (HL) event takes 
$\sim$ 2 (1.5) seconds using an NVIDIA A30 GPU.

For several candidates, source mass posteriors from the GWTC-3 analysis were completely within AMPLFI's training prior, but luminosity distance support only partially overlapped. We analyze these events, but note that AMPLFI's luminosity distance posterior rails against the prior. Still, AMPLFI's sky maps and posterior estimates for source parameters are consistent with the LVK analysis. See \Cref{fig:GW200220_124850-dist-railing,fig:GW200128_022011-dist-railing,fig:GW191222_033537-dist-railing} for examples of such events. A possible solution to this problem would be to train AMPLFI to infer the chirp distance ($d_{\mathrm{chirp}} \propto d_L/\mathcal{M}^{5/6}$) \cite{Roulet_2022, Brady_2008} instead of the luminosity distance. Effectively, this adjusts the distance prior depending on the chirp mass of the source, allowing the recovery of intrinsically louder signals at farther distances. Exploring this is left to future work.

There exist several differences between the GWTC-3 and AMPLFI analyses. First, AMPLFI trains using a distance prior that is uniform, while the GWTC-3 analyses utilize a uniform in co-moving volume distribution. To account for this, we use importance sampling to re-weight AMPLFI's samples to the uniform in co-moving volume distance distribution reported by the GWTC-3 analysis. Second, AMPLFI was trained using the IMRPhenomPV2 waveform, while the LVK posteriors were produced using IMRPhenomXPHM. So, effects of higher order modes are not captured in AMPLFI's results. Third, the AMPLFI result was produced using the low-latency data, while the GWTC-3 result utilized data that includes an update to calibration, and cleaning of certain noise sources. Lastly, AMPLFI does not aim to explicitly model the Gaussian likelihood typically assumed in CBC PE. By training with real data, non-Gaussian statistics in the data is learned.

In general, the posteriors show close agreement. However, there are some instances with noticeable differences. For example, AMPLFI does not show multi-modality in masses exhibited by the GWTC-3 result in GW200129\_065458 (see Figure~\ref{fig:GW200129_065458-hlv}). For this same event, AMPLFI's distance and inclination estimates disagree with the GWTC-3 result. With an SNR of $\sim$ 27, waveform systematics become more important and could explain this discrepancy. For GW191215\_223052 and GW200220\_124850 AMPLFI does not show multi-modality in inclination exhibited by the GWTC-3 result (see~\Cref{fig:GW191215_223052-hlv,fig:GW200220_124850-dist-railing}). Generally, possible sources of the differences between AMPLFI and GWTC-3 results could be the physics of the assumed waveform (e.g. the inclusion of higher order modes in the GWTC-3 result), suboptimal model convergence and expressivity, or the difference in assumed noise statistics (i.e. GWTC-3 PE assumes Gaussian noise, whereas AMPLFI is trained using real data). We note that through importance sampling to a Gaussian likelihood, it is possible to mitigate differences due to suboptimal model convergence and assumed noise statistics \cite{Dax_2023}.

\section{Conclusion and Future Work}
We have presented the performance of the AMPLFI algorithm across several metrics. We have demonstrated AMPLFI's sky localization performance is equivalent to BAYESTAR using injections from an online data replay. We have shown that a single AMPLFI algorithm trained using $\sim 2$ months of real data maintains performance months beyond its training period. Finally, we analyzed real gravitational wave candidates within AMPLFI's training prior and have shown posterior results consistent with the GWTC-3 analyses. 

Still, there remains several avenues for improving the utility of AMPLFI. Broadening the parameter space to include neutron star black hole and and binary neutron star (BNS) mergers will be critical for analyzing more likely candidates of EM emission. For BNS signals that can last for minutes in the detectors' sensitive band, utilizing data compression methods will be important. In addition, exploring parameterizations of the CBC signal parameter space that eliminate known degeneracies could help the learning process \cite{Roulet_2022}. For example, training AMPLFI to infer chirp distance \cite{Brady_2008} instead of luminosity distance could help alleviate the distance prior railing exhibited in AMPLFI's analysis of several real candidates. 

As of August 2025, AMPLFI has been deployed in production for real-time follow-up of CBC candidates detected by the Aframe search algorithm, and has contributed to several public alerts\footnote{\url{https://gracedb.ligo.org/superevents/public/O4c/}}. 

\section{Acknowledgments}
W.B and M.W.C acknowledge support from the National Science Foundation (NSF) with grant numbers
PHY-2010970 and PHY-2117997. E.M, E.K, A.G and
D.C acknowledge support from the NSF under award
PHY-1764464 and PHY-2309200 to the LIGO Laboratory and PHY-2117997 to the A3D3 Institute. E.M,
E.K, A.G and D.C also acknowledge NSF awards PHY1931469 and PHY-1934700. 

This research was undertaken with the support of the
LIGO computational clusters. This material is based upon work supported by NSF's LIGO Laboratory which is a major facility fully funded by the National Science Foundation. This work used NCSA-Delta at U. Illinois
through allocation PHY-240078 from the Advanced Cyberinfrastructure Coordination
Ecosystem: Services and Support (ACCESS) program, which is supported by National
Science Foundation grants \#2138259, \#2138286, \#2138307, \#2137603, and \#2138296. This research
has made use of data or software obtained from the Gravitational Wave Open Science
Center (gwosc.org)  a service of the LIGO Scientific Collaboration, the Virgo
Collaboration, and KAGRA.

\bibliographystyle{apsrev}
\bibliography{references.bib}

\begin{thebibliography}{50}
\expandafter\ifx\csname natexlab\endcsname\relax\def\natexlab#1{#1}\fi
\expandafter\ifx\csname bibnamefont\endcsname\relax
  \def\bibnamefont#1{#1}\fi
\expandafter\ifx\csname bibfnamefont\endcsname\relax
  \def\bibfnamefont#1{#1}\fi
\expandafter\ifx\csname citenamefont\endcsname\relax
  \def\citenamefont#1{#1}\fi
\expandafter\ifx\csname url\endcsname\relax
  \def\url#1{\texttt{#1}}\fi
\expandafter\ifx\csname urlprefix\endcsname\relax\def\urlprefix{URL }\fi
\providecommand{\bibinfo}[2]{#2}
\providecommand{\eprint}[2][]{\url{#2}}

\bibitem[{\citenamefont{Abbott et~al.}(2016)\citenamefont{Abbott, Abbott, Abbott, Abernathy, Acernese, Ackley, Adams, Adams, Addesso, Adhikari et~al.}}]{PhysRevLett.116.061102}
\bibinfo{author}{\bibfnamefont{B.~P.} \bibnamefont{Abbott}}, \bibinfo{author}{\bibfnamefont{R.}~\bibnamefont{Abbott}}, \bibinfo{author}{\bibfnamefont{T.~D.} \bibnamefont{Abbott}}, \bibinfo{author}{\bibfnamefont{M.~R.} \bibnamefont{Abernathy}}, \bibinfo{author}{\bibfnamefont{F.}~\bibnamefont{Acernese}}, \bibinfo{author}{\bibfnamefont{K.}~\bibnamefont{Ackley}}, \bibinfo{author}{\bibfnamefont{C.}~\bibnamefont{Adams}}, \bibinfo{author}{\bibfnamefont{T.}~\bibnamefont{Adams}}, \bibinfo{author}{\bibfnamefont{P.}~\bibnamefont{Addesso}}, \bibinfo{author}{\bibfnamefont{R.~X.} \bibnamefont{Adhikari}}, \bibnamefont{et~al.} (\bibinfo{collaboration}{LIGO Scientific Collaboration and Virgo Collaboration}), \bibinfo{journal}{Phys. Rev. Lett.} \textbf{\bibinfo{volume}{116}}, \bibinfo{pages}{061102} (\bibinfo{year}{2016}), \urlprefix\url{https://link.aps.org/doi/10.1103/PhysRevLett.116.061102}.

\bibitem[{\citenamefont{Abbott et~al.}(2019)\citenamefont{Abbott, Abbott, Abbott, Abraham, Acernese, Ackley, Adams, Adhikari, Adya, Affeldt et~al.}}]{AbEA2018b}
\bibinfo{author}{\bibfnamefont{B.~P.} \bibnamefont{Abbott}}, \bibinfo{author}{\bibfnamefont{R.}~\bibnamefont{Abbott}}, \bibinfo{author}{\bibfnamefont{T.~D.} \bibnamefont{Abbott}}, \bibinfo{author}{\bibfnamefont{S.}~\bibnamefont{Abraham}}, \bibinfo{author}{\bibfnamefont{F.}~\bibnamefont{Acernese}}, \bibinfo{author}{\bibfnamefont{K.}~\bibnamefont{Ackley}}, \bibinfo{author}{\bibfnamefont{C.}~\bibnamefont{Adams}}, \bibinfo{author}{\bibfnamefont{R.~X.} \bibnamefont{Adhikari}}, \bibinfo{author}{\bibfnamefont{V.~B.} \bibnamefont{Adya}}, \bibinfo{author}{\bibfnamefont{C.}~\bibnamefont{Affeldt}}, \bibnamefont{et~al.} (\bibinfo{collaboration}{LIGO Scientific Collaboration and Virgo Collaboration}), \bibinfo{journal}{Phys. Rev. X} \textbf{\bibinfo{volume}{9}}, \bibinfo{pages}{031040} (\bibinfo{year}{2019}), \urlprefix\url{https://link.aps.org/doi/10.1103/PhysRevX.9.031040}.

\bibitem[{\citenamefont{Abbott et~al.}(2024)}]{theligoscientificcollaboration2022gwtc21}
\bibinfo{author}{\bibfnamefont{R.}~\bibnamefont{Abbott}} \bibnamefont{et~al.} (\bibinfo{collaboration}{LIGO Scientific, VIRGO}), \bibinfo{journal}{Phys. Rev. D} \textbf{\bibinfo{volume}{109}}, \bibinfo{pages}{022001} (\bibinfo{year}{2024}), \eprint{2108.01045}.

\bibitem[{\citenamefont{Abbott et~al.}(2023)\citenamefont{Abbott, Abbott, Acernese, Ackley, Adams, Adhikari, Adhikari, Adya, Affeldt, Agarwal et~al.}}]{Abbott_2023}
\bibinfo{author}{\bibfnamefont{R.}~\bibnamefont{Abbott}}, \bibinfo{author}{\bibfnamefont{T.}~\bibnamefont{Abbott}}, \bibinfo{author}{\bibfnamefont{F.}~\bibnamefont{Acernese}}, \bibinfo{author}{\bibfnamefont{K.}~\bibnamefont{Ackley}}, \bibinfo{author}{\bibfnamefont{C.}~\bibnamefont{Adams}}, \bibinfo{author}{\bibfnamefont{N.}~\bibnamefont{Adhikari}}, \bibinfo{author}{\bibfnamefont{R.}~\bibnamefont{Adhikari}}, \bibinfo{author}{\bibfnamefont{V.}~\bibnamefont{Adya}}, \bibinfo{author}{\bibfnamefont{C.}~\bibnamefont{Affeldt}}, \bibinfo{author}{\bibfnamefont{D.}~\bibnamefont{Agarwal}}, \bibnamefont{et~al.}, \bibinfo{journal}{Physical Review X} \textbf{\bibinfo{volume}{13}} (\bibinfo{year}{2023}), ISSN \bibinfo{issn}{2160-3308}, \urlprefix\url{http://dx.doi.org/10.1103/PhysRevX.13.041039}.

\bibitem[{\citenamefont{Abac et~al.}(2025)}]{LIGOScientific:2025slb}
\bibinfo{author}{\bibfnamefont{A.~G.} \bibnamefont{Abac}} \bibnamefont{et~al.} (\bibinfo{collaboration}{LIGO Scientific, VIRGO, KAGRA}) (\bibinfo{year}{2025}), \eprint{2508.18082}.

\bibitem[{\citenamefont{{The LIGO Scientific Collaboration} et~al.}(2015)\citenamefont{{The LIGO Scientific Collaboration}, Aasi, Abbott, Abbott, Abbott, Abernathy, Ackley, Adams, Adams, Addesso et~al.}}]{Aasi_2015}
\bibinfo{author}{\bibnamefont{{The LIGO Scientific Collaboration}}}, \bibinfo{author}{\bibfnamefont{J.}~\bibnamefont{Aasi}}, \bibinfo{author}{\bibfnamefont{B.~P.} \bibnamefont{Abbott}}, \bibinfo{author}{\bibfnamefont{R.}~\bibnamefont{Abbott}}, \bibinfo{author}{\bibfnamefont{T.}~\bibnamefont{Abbott}}, \bibinfo{author}{\bibfnamefont{M.~R.} \bibnamefont{Abernathy}}, \bibinfo{author}{\bibfnamefont{K.}~\bibnamefont{Ackley}}, \bibinfo{author}{\bibfnamefont{C.}~\bibnamefont{Adams}}, \bibinfo{author}{\bibfnamefont{T.}~\bibnamefont{Adams}}, \bibinfo{author}{\bibfnamefont{P.}~\bibnamefont{Addesso}}, \bibnamefont{et~al.}, \bibinfo{journal}{Classical and Quantum Gravity} \textbf{\bibinfo{volume}{32}}, \bibinfo{pages}{074001} (\bibinfo{year}{2015}), \urlprefix\url{https://dx.doi.org/10.1088/0264-9381/32/7/074001}.

\bibitem[{\citenamefont{Acernese et~al.}(2015)}]{VIRGO:2014yos}
\bibinfo{author}{\bibfnamefont{F.}~\bibnamefont{Acernese}} \bibnamefont{et~al.} (\bibinfo{collaboration}{VIRGO}), \bibinfo{journal}{Class. Quant. Grav.} \textbf{\bibinfo{volume}{32}}, \bibinfo{pages}{024001} (\bibinfo{year}{2015}), \eprint{1408.3978}.

\bibitem[{\citenamefont{Akutsu et~al.}(2020)\citenamefont{Akutsu, Ando, Arai, Arai, Araki, Araya, Aritomi, Aso, Bae, Bae et~al.}}]{kagra}
\bibinfo{author}{\bibfnamefont{T.}~\bibnamefont{Akutsu}}, \bibinfo{author}{\bibfnamefont{M.}~\bibnamefont{Ando}}, \bibinfo{author}{\bibfnamefont{K.}~\bibnamefont{Arai}}, \bibinfo{author}{\bibfnamefont{Y.}~\bibnamefont{Arai}}, \bibinfo{author}{\bibfnamefont{S.}~\bibnamefont{Araki}}, \bibinfo{author}{\bibfnamefont{A.}~\bibnamefont{Araya}}, \bibinfo{author}{\bibfnamefont{N.}~\bibnamefont{Aritomi}}, \bibinfo{author}{\bibfnamefont{Y.}~\bibnamefont{Aso}}, \bibinfo{author}{\bibfnamefont{S.}~\bibnamefont{Bae}}, \bibinfo{author}{\bibfnamefont{Y.}~\bibnamefont{Bae}}, \bibnamefont{et~al.}, \bibinfo{journal}{Progress of Theoretical and Experimental Physics} \textbf{\bibinfo{volume}{2021}}, \bibinfo{pages}{05A101} (\bibinfo{year}{2020}), ISSN \bibinfo{issn}{2050-3911}, \eprint{https://academic.oup.com/ptep/article-pdf/2021/5/05A101/37974994/ptaa125.pdf}, \urlprefix\url{https://doi.org/10.1093/ptep/ptaa125}.

\bibitem[{\citenamefont{Abbott et~al.}(2020)\citenamefont{Abbott, Abbott, Abbott, Abraham, Acernese, Ackley, Adams, Adya, Affeldt, Agathos et~al.}}]{observing_scenarios}
\bibinfo{author}{\bibfnamefont{B.~P.} \bibnamefont{Abbott}}, \bibinfo{author}{\bibfnamefont{R.}~\bibnamefont{Abbott}}, \bibinfo{author}{\bibfnamefont{T.~D.} \bibnamefont{Abbott}}, \bibinfo{author}{\bibfnamefont{S.}~\bibnamefont{Abraham}}, \bibinfo{author}{\bibfnamefont{F.}~\bibnamefont{Acernese}}, \bibinfo{author}{\bibfnamefont{K.}~\bibnamefont{Ackley}}, \bibinfo{author}{\bibfnamefont{C.}~\bibnamefont{Adams}}, \bibinfo{author}{\bibfnamefont{V.~B.} \bibnamefont{Adya}}, \bibinfo{author}{\bibfnamefont{C.}~\bibnamefont{Affeldt}}, \bibinfo{author}{\bibfnamefont{M.}~\bibnamefont{Agathos}}, \bibnamefont{et~al.}, \bibinfo{journal}{Living Reviews in Relativity} \textbf{\bibinfo{volume}{23}} (\bibinfo{year}{2020}), ISSN \bibinfo{issn}{1433-8351}, \urlprefix\url{http://dx.doi.org/10.1007/s41114-020-00026-9}.

\bibitem[{\citenamefont{Marx et~al.}(2025)}]{aframe-methods}
\bibinfo{author}{\bibfnamefont{E.}~\bibnamefont{Marx}} \bibnamefont{et~al.}, \bibinfo{journal}{Phys. Rev. D} \textbf{\bibinfo{volume}{111}}, \bibinfo{pages}{042010} (\bibinfo{year}{2025}), \eprint{2403.18661}.

\bibitem[{\citenamefont{Nagarajan and Messenger}(2025)}]{sage}
\bibinfo{author}{\bibfnamefont{N.}~\bibnamefont{Nagarajan}} \bibnamefont{and} \bibinfo{author}{\bibfnamefont{C.}~\bibnamefont{Messenger}}, \bibinfo{journal}{arXiv preprint arXiv:2501.13846}  (\bibinfo{year}{2025}).

\bibitem[{\citenamefont{Nousi et~al.}(2023)\citenamefont{Nousi, Koloniari, Passalis, Iosif, Stergioulas, and Tefas}}]{Nousi_2023}
\bibinfo{author}{\bibfnamefont{P.}~\bibnamefont{Nousi}}, \bibinfo{author}{\bibfnamefont{A.~E.} \bibnamefont{Koloniari}}, \bibinfo{author}{\bibfnamefont{N.}~\bibnamefont{Passalis}}, \bibinfo{author}{\bibfnamefont{P.}~\bibnamefont{Iosif}}, \bibinfo{author}{\bibfnamefont{N.}~\bibnamefont{Stergioulas}}, \bibnamefont{and} \bibinfo{author}{\bibfnamefont{A.}~\bibnamefont{Tefas}}, \bibinfo{journal}{Physical Review D} \textbf{\bibinfo{volume}{108}} (\bibinfo{year}{2023}), ISSN \bibinfo{issn}{2470-0029}, \urlprefix\url{http://dx.doi.org/10.1103/PhysRevD.108.024022}.

\bibitem[{\citenamefont{Singer and Price}(2016{\natexlab{a}})}]{bayestar}
\bibinfo{author}{\bibfnamefont{L.~P.} \bibnamefont{Singer}} \bibnamefont{and} \bibinfo{author}{\bibfnamefont{L.~R.} \bibnamefont{Price}}, \bibinfo{journal}{Physical Review D} \textbf{\bibinfo{volume}{93}} (\bibinfo{year}{2016}{\natexlab{a}}), ISSN \bibinfo{issn}{2470-0029}, \urlprefix\url{http://dx.doi.org/10.1103/PhysRevD.93.024013}.

\bibitem[{\citenamefont{Chatterjee et~al.}(2020)\citenamefont{Chatterjee, Ghosh, Brady, Kapadia, Miller, Nissanke, and Pannarale}}]{Chatterjee:2019avs}
\bibinfo{author}{\bibfnamefont{D.}~\bibnamefont{Chatterjee}}, \bibinfo{author}{\bibfnamefont{S.}~\bibnamefont{Ghosh}}, \bibinfo{author}{\bibfnamefont{P.~R.} \bibnamefont{Brady}}, \bibinfo{author}{\bibfnamefont{S.~J.} \bibnamefont{Kapadia}}, \bibinfo{author}{\bibfnamefont{A.~L.} \bibnamefont{Miller}}, \bibinfo{author}{\bibfnamefont{S.}~\bibnamefont{Nissanke}}, \bibnamefont{and} \bibinfo{author}{\bibfnamefont{F.}~\bibnamefont{Pannarale}}, \bibinfo{journal}{Astrophys. J.} \textbf{\bibinfo{volume}{896}}, \bibinfo{pages}{54} (\bibinfo{year}{2020}), \eprint{1911.00116}.

\bibitem[{\citenamefont{Chatterjee et~al.}(2024)}]{Chatterjee:2024pbj}
\bibinfo{author}{\bibfnamefont{D.}~\bibnamefont{Chatterjee}} \bibnamefont{et~al.}, \bibinfo{journal}{Mach. Learn. Sci. Tech.} \textbf{\bibinfo{volume}{5}}, \bibinfo{pages}{045030} (\bibinfo{year}{2024}), \eprint{2407.19048}.

\bibitem[{\citenamefont{Ashton et~al.}(2019)}]{bilby_paper}
\bibinfo{author}{\bibfnamefont{G.}~\bibnamefont{Ashton}} \bibnamefont{et~al.}, \bibinfo{journal}{Astrophys. J. Suppl.} \textbf{\bibinfo{volume}{241}}, \bibinfo{pages}{27} (\bibinfo{year}{2019}), \eprint{1811.02042}.

\bibitem[{\citenamefont{Veitch et~al.}(2015)\citenamefont{Veitch, Raymond, Farr, Farr, Graff, Vitale, Aylott, Blackburn, Christensen, Coughlin et~al.}}]{lalinference_paper}
\bibinfo{author}{\bibfnamefont{J.}~\bibnamefont{Veitch}}, \bibinfo{author}{\bibfnamefont{V.}~\bibnamefont{Raymond}}, \bibinfo{author}{\bibfnamefont{B.}~\bibnamefont{Farr}}, \bibinfo{author}{\bibfnamefont{W.}~\bibnamefont{Farr}}, \bibinfo{author}{\bibfnamefont{P.}~\bibnamefont{Graff}}, \bibinfo{author}{\bibfnamefont{S.}~\bibnamefont{Vitale}}, \bibinfo{author}{\bibfnamefont{B.}~\bibnamefont{Aylott}}, \bibinfo{author}{\bibfnamefont{K.}~\bibnamefont{Blackburn}}, \bibinfo{author}{\bibfnamefont{N.}~\bibnamefont{Christensen}}, \bibinfo{author}{\bibfnamefont{M.}~\bibnamefont{Coughlin}}, \bibnamefont{et~al.}, \bibinfo{journal}{Physical Review D} \textbf{\bibinfo{volume}{91}} (\bibinfo{year}{2015}), ISSN \bibinfo{issn}{1550-2368}, \urlprefix\url{http://dx.doi.org/10.1103/PhysRevD.91.042003}.

\bibitem[{\citenamefont{Canizares et~al.}(2015)\citenamefont{Canizares, Field, Gair, Raymond, Smith, and Tiglio}}]{Canizares:2014fya}
\bibinfo{author}{\bibfnamefont{P.}~\bibnamefont{Canizares}}, \bibinfo{author}{\bibfnamefont{S.~E.} \bibnamefont{Field}}, \bibinfo{author}{\bibfnamefont{J.}~\bibnamefont{Gair}}, \bibinfo{author}{\bibfnamefont{V.}~\bibnamefont{Raymond}}, \bibinfo{author}{\bibfnamefont{R.}~\bibnamefont{Smith}}, \bibnamefont{and} \bibinfo{author}{\bibfnamefont{M.}~\bibnamefont{Tiglio}}, \bibinfo{journal}{Phys. Rev. Lett.} \textbf{\bibinfo{volume}{114}}, \bibinfo{pages}{071104} (\bibinfo{year}{2015}), \eprint{1404.6284}.

\bibitem[{\citenamefont{Morisaki et~al.}(2023)\citenamefont{Morisaki, Smith, Tsukada, Sachdev, Stevenson, Talbot, and Zimmerman}}]{Morisaki:2023kuq}
\bibinfo{author}{\bibfnamefont{S.}~\bibnamefont{Morisaki}}, \bibinfo{author}{\bibfnamefont{R.}~\bibnamefont{Smith}}, \bibinfo{author}{\bibfnamefont{L.}~\bibnamefont{Tsukada}}, \bibinfo{author}{\bibfnamefont{S.}~\bibnamefont{Sachdev}}, \bibinfo{author}{\bibfnamefont{S.}~\bibnamefont{Stevenson}}, \bibinfo{author}{\bibfnamefont{C.}~\bibnamefont{Talbot}}, \bibnamefont{and} \bibinfo{author}{\bibfnamefont{A.}~\bibnamefont{Zimmerman}} (\bibinfo{year}{2023}), \eprint{2307.13380}.

\bibitem[{\citenamefont{Morisaki and Raymond}(2020)}]{Morisaki:2020oqk}
\bibinfo{author}{\bibfnamefont{S.}~\bibnamefont{Morisaki}} \bibnamefont{and} \bibinfo{author}{\bibfnamefont{V.}~\bibnamefont{Raymond}}, \bibinfo{journal}{Phys. Rev. D} \textbf{\bibinfo{volume}{102}}, \bibinfo{pages}{104020} (\bibinfo{year}{2020}), \eprint{2007.09108}.

\bibitem[{\citenamefont{Singer and Price}(2016{\natexlab{b}})}]{SiPr2016}
\bibinfo{author}{\bibfnamefont{L.~P.} \bibnamefont{Singer}} \bibnamefont{and} \bibinfo{author}{\bibfnamefont{L.~R.} \bibnamefont{Price}}, \bibinfo{journal}{Phys. Rev. D} \textbf{\bibinfo{volume}{93}}, \bibinfo{pages}{024013} (\bibinfo{year}{2016}{\natexlab{b}}), \urlprefix\url{https://link.aps.org/doi/10.1103/PhysRevD.93.024013}.

\bibitem[{\citenamefont{Papamakarios et~al.}(2021)\citenamefont{Papamakarios, Nalisnick, Rezende, Mohamed, and Lakshminarayanan}}]{papamakarios2021normalizing}
\bibinfo{author}{\bibfnamefont{G.}~\bibnamefont{Papamakarios}}, \bibinfo{author}{\bibfnamefont{E.}~\bibnamefont{Nalisnick}}, \bibinfo{author}{\bibfnamefont{D.~J.} \bibnamefont{Rezende}}, \bibinfo{author}{\bibfnamefont{S.}~\bibnamefont{Mohamed}}, \bibnamefont{and} \bibinfo{author}{\bibfnamefont{B.}~\bibnamefont{Lakshminarayanan}}, \bibinfo{journal}{Journal of Machine Learning Research} \textbf{\bibinfo{volume}{22}}, \bibinfo{pages}{1} (\bibinfo{year}{2021}).

\bibitem[{\citenamefont{Dax et~al.}(2021)\citenamefont{Dax, Green, Gair, Macke, Buonanno, and Schölkopf}}]{Dax_2021}
\bibinfo{author}{\bibfnamefont{M.}~\bibnamefont{Dax}}, \bibinfo{author}{\bibfnamefont{S.~R.} \bibnamefont{Green}}, \bibinfo{author}{\bibfnamefont{J.}~\bibnamefont{Gair}}, \bibinfo{author}{\bibfnamefont{J.~H.} \bibnamefont{Macke}}, \bibinfo{author}{\bibfnamefont{A.}~\bibnamefont{Buonanno}}, \bibnamefont{and} \bibinfo{author}{\bibfnamefont{B.}~\bibnamefont{Schölkopf}}, \bibinfo{journal}{Physical Review Letters} \textbf{\bibinfo{volume}{127}} (\bibinfo{year}{2021}), ISSN \bibinfo{issn}{1079-7114}, \urlprefix\url{http://dx.doi.org/10.1103/PhysRevLett.127.241103}.

\bibitem[{\citenamefont{Dax et~al.}(2025)\citenamefont{Dax, Green, Gair, Gupte, Pürrer, Raymond, Wildberger, Macke, Buonanno, and Schölkopf}}]{Dax_2025}
\bibinfo{author}{\bibfnamefont{M.}~\bibnamefont{Dax}}, \bibinfo{author}{\bibfnamefont{S.~R.} \bibnamefont{Green}}, \bibinfo{author}{\bibfnamefont{J.}~\bibnamefont{Gair}}, \bibinfo{author}{\bibfnamefont{N.}~\bibnamefont{Gupte}}, \bibinfo{author}{\bibfnamefont{M.}~\bibnamefont{Pürrer}}, \bibinfo{author}{\bibfnamefont{V.}~\bibnamefont{Raymond}}, \bibinfo{author}{\bibfnamefont{J.}~\bibnamefont{Wildberger}}, \bibinfo{author}{\bibfnamefont{J.~H.} \bibnamefont{Macke}}, \bibinfo{author}{\bibfnamefont{A.}~\bibnamefont{Buonanno}}, \bibnamefont{and} \bibinfo{author}{\bibfnamefont{B.}~\bibnamefont{Schölkopf}}, \bibinfo{journal}{Nature} \textbf{\bibinfo{volume}{639}}, \bibinfo{pages}{49–53} (\bibinfo{year}{2025}), ISSN \bibinfo{issn}{1476-4687}, \urlprefix\url{http://dx.doi.org/10.1038/s41586-025-08593-z}.

\bibitem[{\citenamefont{Williams et~al.}(2021)\citenamefont{Williams, Veitch, and Messenger}}]{Williams:2021qyt}
\bibinfo{author}{\bibfnamefont{M.~J.} \bibnamefont{Williams}}, \bibinfo{author}{\bibfnamefont{J.}~\bibnamefont{Veitch}}, \bibnamefont{and} \bibinfo{author}{\bibfnamefont{C.}~\bibnamefont{Messenger}}, \bibinfo{journal}{Phys. Rev. D} \textbf{\bibinfo{volume}{103}}, \bibinfo{pages}{103006} (\bibinfo{year}{2021}), \eprint{2102.11056}.

\bibitem[{\citenamefont{De~Santi et~al.}(2024)\citenamefont{De~Santi, Razzano, Fidecaro, Muccillo, Papalini, and Patricelli}}]{PhysRevD.109.102004}
\bibinfo{author}{\bibfnamefont{F.}~\bibnamefont{De~Santi}}, \bibinfo{author}{\bibfnamefont{M.}~\bibnamefont{Razzano}}, \bibinfo{author}{\bibfnamefont{F.}~\bibnamefont{Fidecaro}}, \bibinfo{author}{\bibfnamefont{L.}~\bibnamefont{Muccillo}}, \bibinfo{author}{\bibfnamefont{L.}~\bibnamefont{Papalini}}, \bibnamefont{and} \bibinfo{author}{\bibfnamefont{B.}~\bibnamefont{Patricelli}}, \bibinfo{journal}{Phys. Rev. D} \textbf{\bibinfo{volume}{109}}, \bibinfo{pages}{102004} (\bibinfo{year}{2024}), \urlprefix\url{https://link.aps.org/doi/10.1103/PhysRevD.109.102004}.

\bibitem[{\citenamefont{Bellm et~al.}(2018)\citenamefont{Bellm, Kulkarni, Graham, Dekany, Smith, Riddle, Masci, Helou, Prince, Adams et~al.}}]{Bellm_2018}
\bibinfo{author}{\bibfnamefont{E.~C.} \bibnamefont{Bellm}}, \bibinfo{author}{\bibfnamefont{S.~R.} \bibnamefont{Kulkarni}}, \bibinfo{author}{\bibfnamefont{M.~J.} \bibnamefont{Graham}}, \bibinfo{author}{\bibfnamefont{R.}~\bibnamefont{Dekany}}, \bibinfo{author}{\bibfnamefont{R.~M.} \bibnamefont{Smith}}, \bibinfo{author}{\bibfnamefont{R.}~\bibnamefont{Riddle}}, \bibinfo{author}{\bibfnamefont{F.~J.} \bibnamefont{Masci}}, \bibinfo{author}{\bibfnamefont{G.}~\bibnamefont{Helou}}, \bibinfo{author}{\bibfnamefont{T.~A.} \bibnamefont{Prince}}, \bibinfo{author}{\bibfnamefont{S.~M.} \bibnamefont{Adams}}, \bibnamefont{et~al.}, \bibinfo{journal}{Publications of the Astronomical Society of the Pacific} \textbf{\bibinfo{volume}{131}}, \bibinfo{pages}{018002} (\bibinfo{year}{2018}), ISSN \bibinfo{issn}{1538-3873}, \urlprefix\url{http://dx.doi.org/10.1088/1538-3873/aaecbe}.

\bibitem[{\citenamefont{Khan et~al.}(2016)\citenamefont{Khan, Husa, Hannam, Ohme, P\"urrer, Forteza, and Boh\'e}}]{phenomD}
\bibinfo{author}{\bibfnamefont{S.}~\bibnamefont{Khan}}, \bibinfo{author}{\bibfnamefont{S.}~\bibnamefont{Husa}}, \bibinfo{author}{\bibfnamefont{M.}~\bibnamefont{Hannam}}, \bibinfo{author}{\bibfnamefont{F.}~\bibnamefont{Ohme}}, \bibinfo{author}{\bibfnamefont{M.}~\bibnamefont{P\"urrer}}, \bibinfo{author}{\bibfnamefont{X.~J.} \bibnamefont{Forteza}}, \bibnamefont{and} \bibinfo{author}{\bibfnamefont{A.}~\bibnamefont{Boh\'e}}, \bibinfo{journal}{Phys. Rev. D} \textbf{\bibinfo{volume}{93}}, \bibinfo{pages}{044007} (\bibinfo{year}{2016}), \urlprefix\url{https://link.aps.org/doi/10.1103/PhysRevD.93.044007}.

\bibitem[{\citenamefont{He et~al.}(2016)\citenamefont{He, Zhang, Ren, and Sun}}]{resnet}
\bibinfo{author}{\bibfnamefont{K.}~\bibnamefont{He}}, \bibinfo{author}{\bibfnamefont{X.}~\bibnamefont{Zhang}}, \bibinfo{author}{\bibfnamefont{S.}~\bibnamefont{Ren}}, \bibnamefont{and} \bibinfo{author}{\bibfnamefont{J.}~\bibnamefont{Sun}}, in \emph{\bibinfo{booktitle}{2016 IEEE Conference on Computer Vision and Pattern Recognition (CVPR)}} (\bibinfo{year}{2016}), pp. \bibinfo{pages}{770--778}.

\bibitem[{\citenamefont{Bhardwaj et~al.}(2023)\citenamefont{Bhardwaj, Alvey, Miller, Nissanke, and Weniger}}]{Bhardwaj:2023xph}
\bibinfo{author}{\bibfnamefont{U.}~\bibnamefont{Bhardwaj}}, \bibinfo{author}{\bibfnamefont{J.}~\bibnamefont{Alvey}}, \bibinfo{author}{\bibfnamefont{B.~K.} \bibnamefont{Miller}}, \bibinfo{author}{\bibfnamefont{S.}~\bibnamefont{Nissanke}}, \bibnamefont{and} \bibinfo{author}{\bibfnamefont{C.}~\bibnamefont{Weniger}}, \bibinfo{journal}{Phys. Rev. D} \textbf{\bibinfo{volume}{108}}, \bibinfo{pages}{042004} (\bibinfo{year}{2023}), \eprint{2304.02035}.

\bibitem[{\citenamefont{Sun et~al.}(2024)\citenamefont{Sun, Xiong, Jin, Wang, Zhang, and Zhang}}]{Sun:2023vlq}
\bibinfo{author}{\bibfnamefont{T.-Y.} \bibnamefont{Sun}}, \bibinfo{author}{\bibfnamefont{C.-Y.} \bibnamefont{Xiong}}, \bibinfo{author}{\bibfnamefont{S.-J.} \bibnamefont{Jin}}, \bibinfo{author}{\bibfnamefont{Y.-X.} \bibnamefont{Wang}}, \bibinfo{author}{\bibfnamefont{J.-F.} \bibnamefont{Zhang}}, \bibnamefont{and} \bibinfo{author}{\bibfnamefont{X.}~\bibnamefont{Zhang}}, \bibinfo{journal}{Chin. Phys. C} \textbf{\bibinfo{volume}{48}}, \bibinfo{pages}{045108} (\bibinfo{year}{2024}), \eprint{2312.08122}.

\bibitem[{\citenamefont{Dinh et~al.}(2017)\citenamefont{Dinh, Sohl-Dickstein, and Bengio}}]{realnvp}
\bibinfo{author}{\bibfnamefont{L.}~\bibnamefont{Dinh}}, \bibinfo{author}{\bibfnamefont{J.}~\bibnamefont{Sohl-Dickstein}}, \bibnamefont{and} \bibinfo{author}{\bibfnamefont{S.}~\bibnamefont{Bengio}}, \emph{\bibinfo{title}{Density estimation using real nvp}} (\bibinfo{year}{2017}), \eprint{1605.08803}, \urlprefix\url{https://arxiv.org/abs/1605.08803}.

\bibitem[{\citenamefont{Durkan et~al.}(2019)\citenamefont{Durkan, Bekasov, Murray, and Papamakarios}}]{spline}
\bibinfo{author}{\bibfnamefont{C.}~\bibnamefont{Durkan}}, \bibinfo{author}{\bibfnamefont{A.}~\bibnamefont{Bekasov}}, \bibinfo{author}{\bibfnamefont{I.}~\bibnamefont{Murray}}, \bibnamefont{and} \bibinfo{author}{\bibfnamefont{G.}~\bibnamefont{Papamakarios}}, \bibinfo{journal}{Advances in neural information processing systems} \textbf{\bibinfo{volume}{32}} (\bibinfo{year}{2019}).

\bibitem[{\citenamefont{Rozet et~al.}(2022)}]{rozet2022zuko}
\bibinfo{author}{\bibfnamefont{F.}~\bibnamefont{Rozet}} \bibnamefont{et~al.}, \emph{\bibinfo{title}{{Zuko}: Normalizing flows in pytorch}} (\bibinfo{year}{2022}), \urlprefix\url{https://pypi.org/project/zuko}.

\bibitem[{\citenamefont{Chaudhary et~al.}(2024)}]{Chaudhary:2023vec}
\bibinfo{author}{\bibfnamefont{S.~S.} \bibnamefont{Chaudhary}} \bibnamefont{et~al.}, \bibinfo{journal}{Proc. Nat. Acad. Sci.} \textbf{\bibinfo{volume}{121}}, \bibinfo{pages}{e2316474121} (\bibinfo{year}{2024}), \eprint{2308.04545}.

\bibitem[{\citenamefont{Zonca et~al.}(2019)\citenamefont{Zonca, Singer, Lenz, Reinecke, Rosset, Hivon, and Gorski}}]{Zonca2019}
\bibinfo{author}{\bibfnamefont{A.}~\bibnamefont{Zonca}}, \bibinfo{author}{\bibfnamefont{L.}~\bibnamefont{Singer}}, \bibinfo{author}{\bibfnamefont{D.}~\bibnamefont{Lenz}}, \bibinfo{author}{\bibfnamefont{M.}~\bibnamefont{Reinecke}}, \bibinfo{author}{\bibfnamefont{C.}~\bibnamefont{Rosset}}, \bibinfo{author}{\bibfnamefont{E.}~\bibnamefont{Hivon}}, \bibnamefont{and} \bibinfo{author}{\bibfnamefont{K.}~\bibnamefont{Gorski}}, \bibinfo{journal}{Journal of Open Source Software} \textbf{\bibinfo{volume}{4}}, \bibinfo{pages}{1298} (\bibinfo{year}{2019}), \urlprefix\url{https://doi.org/10.21105/joss.01298}.

\bibitem[{\citenamefont{{Singer et al.}}(2016)}]{Si2016}
\bibinfo{author}{\bibnamefont{{Singer et al.}}}, \bibinfo{journal}{The Astrophysical Journal Letters} \textbf{\bibinfo{volume}{829}}, \bibinfo{pages}{L15} (\bibinfo{year}{2016}), \urlprefix\url{http://stacks.iop.org/2041-8205/829/i=1/a=L15}.

\bibitem[{\citenamefont{Andreoni et~al.}(2024)\citenamefont{Andreoni, Margutti, Banovetz, Greenstreet, Hebert, Lister, Palmese, Piranomonte, Smartt, Smith et~al.}}]{andreoni2024rubin2024envisioningvera}
\bibinfo{author}{\bibfnamefont{I.}~\bibnamefont{Andreoni}}, \bibinfo{author}{\bibfnamefont{R.}~\bibnamefont{Margutti}}, \bibinfo{author}{\bibfnamefont{J.}~\bibnamefont{Banovetz}}, \bibinfo{author}{\bibfnamefont{S.}~\bibnamefont{Greenstreet}}, \bibinfo{author}{\bibfnamefont{C.-A.} \bibnamefont{Hebert}}, \bibinfo{author}{\bibfnamefont{T.}~\bibnamefont{Lister}}, \bibinfo{author}{\bibfnamefont{A.}~\bibnamefont{Palmese}}, \bibinfo{author}{\bibfnamefont{S.}~\bibnamefont{Piranomonte}}, \bibinfo{author}{\bibfnamefont{S.}~\bibnamefont{Smartt}}, \bibinfo{author}{\bibfnamefont{G.~P.} \bibnamefont{Smith}}, \bibnamefont{et~al.}, \bibinfo{journal}{arXiv preprint arXiv:2411.04793}  (\bibinfo{year}{2024}).

\bibitem[{\citenamefont{Singer et~al.}(2016)\citenamefont{Singer, Chen, Holz, Farr, Price, Raymond, Cenko, Gehrels, Cannizzo, Kasliwal et~al.}}]{Singer_2016}
\bibinfo{author}{\bibfnamefont{L.~P.} \bibnamefont{Singer}}, \bibinfo{author}{\bibfnamefont{H.-Y.} \bibnamefont{Chen}}, \bibinfo{author}{\bibfnamefont{D.~E.} \bibnamefont{Holz}}, \bibinfo{author}{\bibfnamefont{W.~M.} \bibnamefont{Farr}}, \bibinfo{author}{\bibfnamefont{L.~R.} \bibnamefont{Price}}, \bibinfo{author}{\bibfnamefont{V.}~\bibnamefont{Raymond}}, \bibinfo{author}{\bibfnamefont{S.~B.} \bibnamefont{Cenko}}, \bibinfo{author}{\bibfnamefont{N.}~\bibnamefont{Gehrels}}, \bibinfo{author}{\bibfnamefont{J.}~\bibnamefont{Cannizzo}}, \bibinfo{author}{\bibfnamefont{M.~M.} \bibnamefont{Kasliwal}}, \bibnamefont{et~al.}, \bibinfo{journal}{The Astrophysical Journal Supplement Series} \textbf{\bibinfo{volume}{226}}, \bibinfo{pages}{10} (\bibinfo{year}{2016}), \urlprefix\url{https://dx.doi.org/10.3847/0067-0049/226/1/10}.

\bibitem[{\citenamefont{Fernique et~al.}(2014)\citenamefont{Fernique, Boch, Donaldson, Durand, O’Mullane, Reinecke, and Taylor}}]{Fernique}
\bibinfo{author}{\bibfnamefont{P.}~\bibnamefont{Fernique}}, \bibinfo{author}{\bibfnamefont{T.}~\bibnamefont{Boch}}, \bibinfo{author}{\bibfnamefont{T.}~\bibnamefont{Donaldson}}, \bibinfo{author}{\bibfnamefont{D.}~\bibnamefont{Durand}}, \bibinfo{author}{\bibfnamefont{W.}~\bibnamefont{O’Mullane}}, \bibinfo{author}{\bibfnamefont{M.}~\bibnamefont{Reinecke}}, \bibnamefont{and} \bibinfo{author}{\bibfnamefont{M.}~\bibnamefont{Taylor}} (\bibinfo{year}{2014}), \urlprefix\url{http://dx.doi.org/10.5479/ADS/bib/2014ivoa.spec.0602F}.

\bibitem[{\citenamefont{Essick et~al.}(2015)\citenamefont{Essick, Vitale, Katsavounidis, Vedovato, and Klimenko}}]{Essick:2014wwa}
\bibinfo{author}{\bibfnamefont{R.}~\bibnamefont{Essick}}, \bibinfo{author}{\bibfnamefont{S.}~\bibnamefont{Vitale}}, \bibinfo{author}{\bibfnamefont{E.}~\bibnamefont{Katsavounidis}}, \bibinfo{author}{\bibfnamefont{G.}~\bibnamefont{Vedovato}}, \bibnamefont{and} \bibinfo{author}{\bibfnamefont{S.}~\bibnamefont{Klimenko}}, \bibinfo{journal}{Astrophys. J.} \textbf{\bibinfo{volume}{800}}, \bibinfo{pages}{81} (\bibinfo{year}{2015}), \eprint{1409.2435}.

\bibitem[{\citenamefont{Nitz et~al.}(2018)\citenamefont{Nitz, Dal~Canton, Davis, and Reyes}}]{Nitz_2018}
\bibinfo{author}{\bibfnamefont{A.~H.} \bibnamefont{Nitz}}, \bibinfo{author}{\bibfnamefont{T.}~\bibnamefont{Dal~Canton}}, \bibinfo{author}{\bibfnamefont{D.}~\bibnamefont{Davis}}, \bibnamefont{and} \bibinfo{author}{\bibfnamefont{S.}~\bibnamefont{Reyes}}, \bibinfo{journal}{Physical Review D} \textbf{\bibinfo{volume}{98}} (\bibinfo{year}{2018}), ISSN \bibinfo{issn}{2470-0029}, \urlprefix\url{http://dx.doi.org/10.1103/PhysRevD.98.024050}.

\bibitem[{\citenamefont{Duverne et~al.}(2024)\citenamefont{Duverne, Hoang, Dal~Canton, Antier, Arnaud, Hello, and Pannarale}}]{Duverne_2024}
\bibinfo{author}{\bibfnamefont{P.-A.} \bibnamefont{Duverne}}, \bibinfo{author}{\bibfnamefont{S.}~\bibnamefont{Hoang}}, \bibinfo{author}{\bibfnamefont{T.}~\bibnamefont{Dal~Canton}}, \bibinfo{author}{\bibfnamefont{S.}~\bibnamefont{Antier}}, \bibinfo{author}{\bibfnamefont{N.}~\bibnamefont{Arnaud}}, \bibinfo{author}{\bibfnamefont{P.}~\bibnamefont{Hello}}, \bibnamefont{and} \bibinfo{author}{\bibfnamefont{F.}~\bibnamefont{Pannarale}}, \bibinfo{journal}{Physical Review D} \textbf{\bibinfo{volume}{110}} (\bibinfo{year}{2024}), ISSN \bibinfo{issn}{2470-0029}, \urlprefix\url{http://dx.doi.org/10.1103/PhysRevD.110.102002}.

\bibitem[{\citenamefont{Ewing et~al.}(2024)}]{Ewing:2023qqe}
\bibinfo{author}{\bibfnamefont{B.}~\bibnamefont{Ewing}} \bibnamefont{et~al.}, \bibinfo{journal}{Phys. Rev. D} \textbf{\bibinfo{volume}{109}}, \bibinfo{pages}{042008} (\bibinfo{year}{2024}), \eprint{2305.05625}.

\bibitem[{\citenamefont{Davis et~al.}(2021)\citenamefont{Davis, Areeda, Berger, Bruntz, Effler, Essick, Fisher, Godwin, Goetz, Helmling-Cornell et~al.}}]{Davis_2021}
\bibinfo{author}{\bibfnamefont{D.}~\bibnamefont{Davis}}, \bibinfo{author}{\bibfnamefont{J.~S.} \bibnamefont{Areeda}}, \bibinfo{author}{\bibfnamefont{B.~K.} \bibnamefont{Berger}}, \bibinfo{author}{\bibfnamefont{R.}~\bibnamefont{Bruntz}}, \bibinfo{author}{\bibfnamefont{A.}~\bibnamefont{Effler}}, \bibinfo{author}{\bibfnamefont{R.~C.} \bibnamefont{Essick}}, \bibinfo{author}{\bibfnamefont{R.~P.} \bibnamefont{Fisher}}, \bibinfo{author}{\bibfnamefont{P.}~\bibnamefont{Godwin}}, \bibinfo{author}{\bibfnamefont{E.}~\bibnamefont{Goetz}}, \bibinfo{author}{\bibfnamefont{A.~F.} \bibnamefont{Helmling-Cornell}}, \bibnamefont{et~al.}, \bibinfo{journal}{Classical and Quantum Gravity} \textbf{\bibinfo{volume}{38}}, \bibinfo{pages}{135014} (\bibinfo{year}{2021}), \urlprefix\url{https://dx.doi.org/10.1088/1361-6382/abfd85}.

\bibitem[{\citenamefont{Soni et~al.}(2025)}]{LIGO:2024kkz}
\bibinfo{author}{\bibfnamefont{S.}~\bibnamefont{Soni}} \bibnamefont{et~al.} (\bibinfo{collaboration}{LIGO}), \bibinfo{journal}{Class. Quant. Grav.} \textbf{\bibinfo{volume}{42}}, \bibinfo{pages}{085016} (\bibinfo{year}{2025}), \eprint{2409.02831}.

\bibitem[{\citenamefont{Wildberger et~al.}(2023)\citenamefont{Wildberger, Dax, Green, Gair, Pürrer, Macke, Buonanno, and Schölkopf}}]{Wildberger_2023}
\bibinfo{author}{\bibfnamefont{J.}~\bibnamefont{Wildberger}}, \bibinfo{author}{\bibfnamefont{M.}~\bibnamefont{Dax}}, \bibinfo{author}{\bibfnamefont{S.~R.} \bibnamefont{Green}}, \bibinfo{author}{\bibfnamefont{J.}~\bibnamefont{Gair}}, \bibinfo{author}{\bibfnamefont{M.}~\bibnamefont{Pürrer}}, \bibinfo{author}{\bibfnamefont{J.~H.} \bibnamefont{Macke}}, \bibinfo{author}{\bibfnamefont{A.}~\bibnamefont{Buonanno}}, \bibnamefont{and} \bibinfo{author}{\bibfnamefont{B.}~\bibnamefont{Schölkopf}}, \bibinfo{journal}{Physical Review D} \textbf{\bibinfo{volume}{107}} (\bibinfo{year}{2023}), ISSN \bibinfo{issn}{2470-0029}, \urlprefix\url{http://dx.doi.org/10.1103/PhysRevD.107.084046}.

\bibitem[{\citenamefont{Roulet et~al.}(2022)\citenamefont{Roulet, Olsen, Mushkin, Islam, Venumadhav, Zackay, and Zaldarriaga}}]{Roulet_2022}
\bibinfo{author}{\bibfnamefont{J.}~\bibnamefont{Roulet}}, \bibinfo{author}{\bibfnamefont{S.}~\bibnamefont{Olsen}}, \bibinfo{author}{\bibfnamefont{J.}~\bibnamefont{Mushkin}}, \bibinfo{author}{\bibfnamefont{T.}~\bibnamefont{Islam}}, \bibinfo{author}{\bibfnamefont{T.}~\bibnamefont{Venumadhav}}, \bibinfo{author}{\bibfnamefont{B.}~\bibnamefont{Zackay}}, \bibnamefont{and} \bibinfo{author}{\bibfnamefont{M.}~\bibnamefont{Zaldarriaga}}, \bibinfo{journal}{Physical Review D} \textbf{\bibinfo{volume}{106}} (\bibinfo{year}{2022}), ISSN \bibinfo{issn}{2470-0029}, \urlprefix\url{http://dx.doi.org/10.1103/PhysRevD.106.123015}.

\bibitem[{\citenamefont{Brady and Fairhurst}(2008)}]{Brady_2008}
\bibinfo{author}{\bibfnamefont{P.~R.} \bibnamefont{Brady}} \bibnamefont{and} \bibinfo{author}{\bibfnamefont{S.}~\bibnamefont{Fairhurst}}, \bibinfo{journal}{Classical and Quantum Gravity} \textbf{\bibinfo{volume}{25}}, \bibinfo{pages}{105002} (\bibinfo{year}{2008}), ISSN \bibinfo{issn}{1361-6382}, \urlprefix\url{http://dx.doi.org/10.1088/0264-9381/25/10/105002}.

\bibitem[{\citenamefont{Dax et~al.}(2023)\citenamefont{Dax, Green, Gair, Pürrer, Wildberger, Macke, Buonanno, and Schölkopf}}]{Dax_2023}
\bibinfo{author}{\bibfnamefont{M.}~\bibnamefont{Dax}}, \bibinfo{author}{\bibfnamefont{S.~R.} \bibnamefont{Green}}, \bibinfo{author}{\bibfnamefont{J.}~\bibnamefont{Gair}}, \bibinfo{author}{\bibfnamefont{M.}~\bibnamefont{Pürrer}}, \bibinfo{author}{\bibfnamefont{J.}~\bibnamefont{Wildberger}}, \bibinfo{author}{\bibfnamefont{J.~H.} \bibnamefont{Macke}}, \bibinfo{author}{\bibfnamefont{A.}~\bibnamefont{Buonanno}}, \bibnamefont{and} \bibinfo{author}{\bibfnamefont{B.}~\bibnamefont{Schölkopf}}, \bibinfo{journal}{Physical Review Letters} \textbf{\bibinfo{volume}{130}} (\bibinfo{year}{2023}), ISSN \bibinfo{issn}{1079-7114}, \urlprefix\url{http://dx.doi.org/10.1103/PhysRevLett.130.171403}.

\end{thebibliography}

\appendix

\begin{figure*}
    \centering
    \includegraphics[scale=0.4]{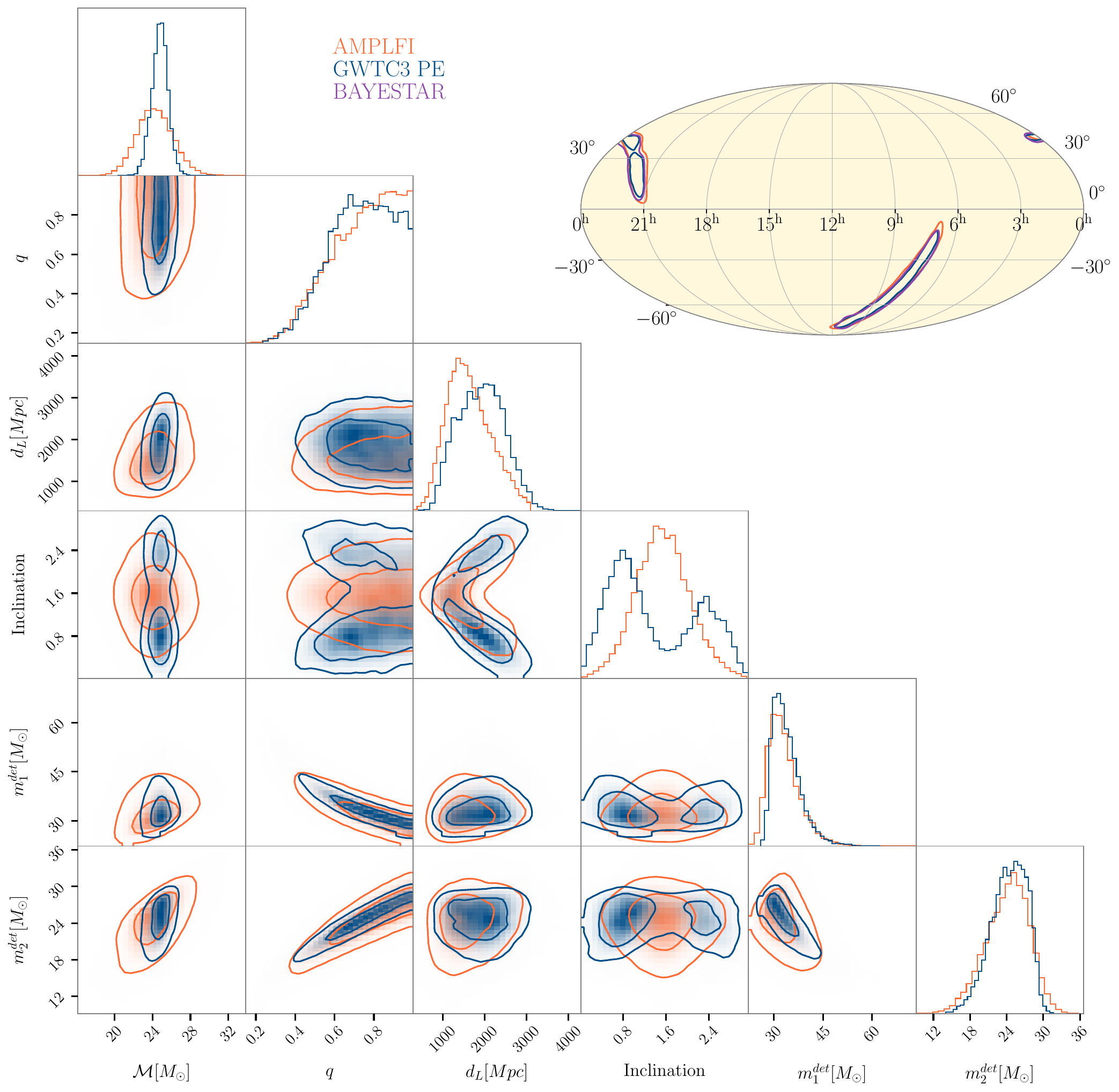}
    \caption{Posterior comparison between AMPLFI and GWTC-3 result for GW191215\_223052. Hanford, Livingston and Virgo data is analyzed. The BAYESTAR sky map produced in low-latency is plotted in purple. Sky map contours correspond to 90\% confidence intervals.}
    \label{fig:GW191215_223052-hlv}
\end{figure*}

\begin{figure*}
    \centering
    \includegraphics[scale=0.4]{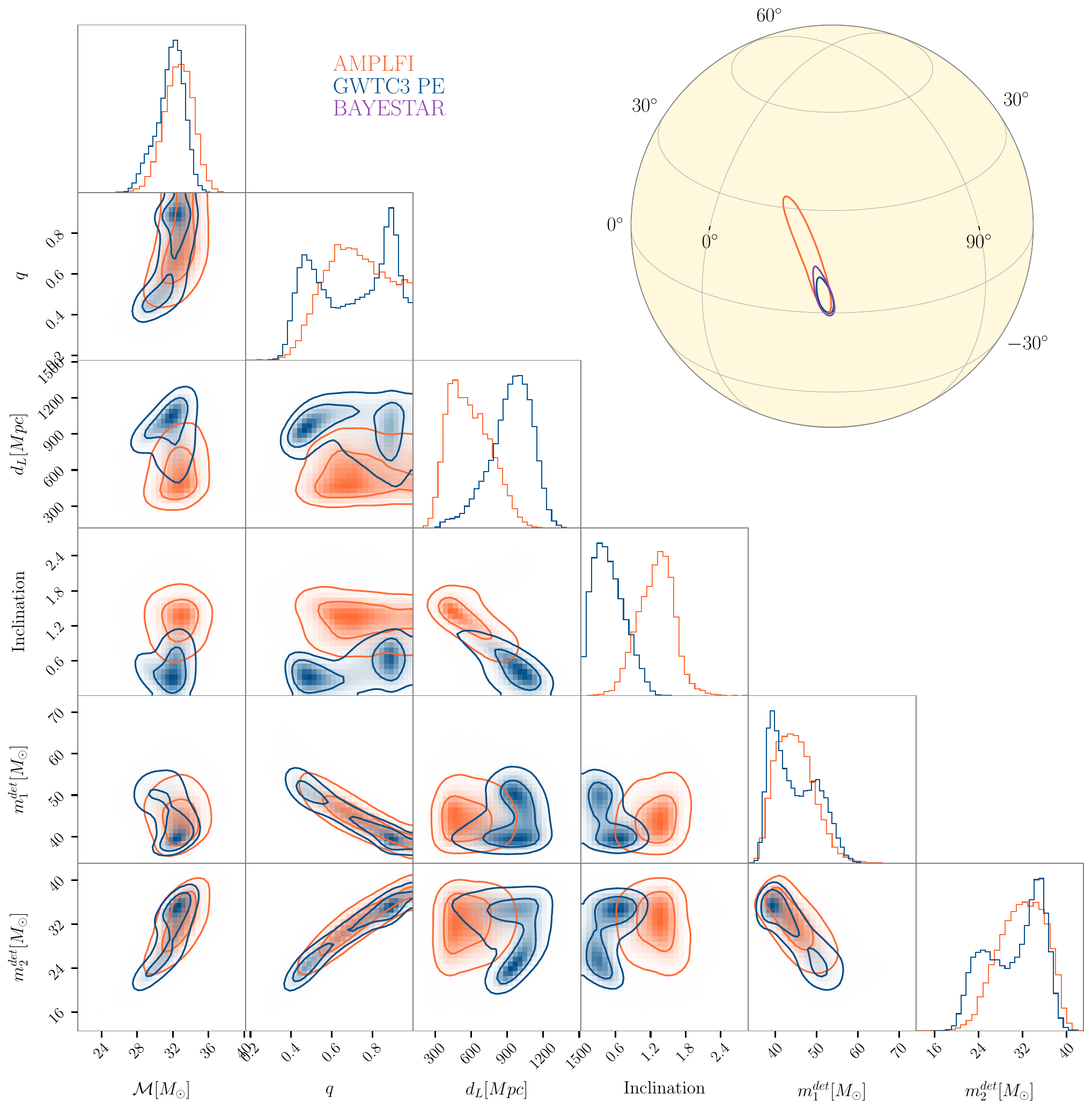}
    \caption{Posterior comparison between AMPLFI and GWTC-3 result for GW200129\_065458. Hanford, Livingston and Virgo data is analyzed. The BAYESTAR sky map produced in low-latency is plotted in purple. Sky map contours correspond to 90\% confidence intervals. For this event, bi-modalities in $m_1$ and $m_2$ are not captured by AMPLFI. In addition, AMPLFI's distance and inclination posteriors disagree with the GWTC-3 result.}
    \label{fig:GW200129_065458-hlv}
\end{figure*}

\begin{figure*}
    \centering
    \includegraphics[scale=0.4]{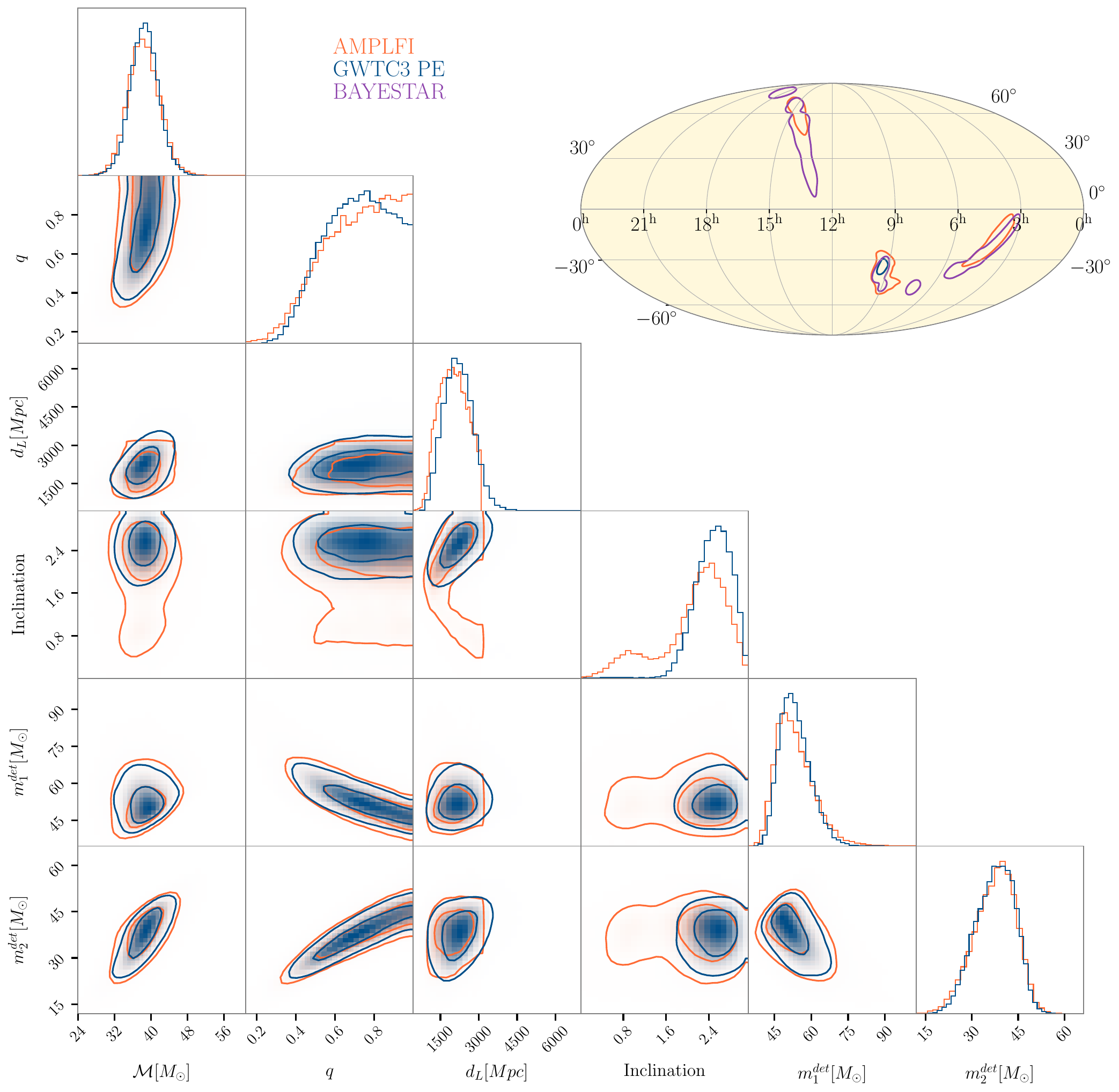}
    \caption{Posterior comparison between AMPLFI and GWTC-3 result for GW200208\_130117. Hanford, Livingston and Virgo data is analyzed. The BAYESTAR sky map produced in low-latency is plotted in purple. Sky map contours correspond to 90\% confidence intervals.}
    \label{fig:GW200208_130117-hlv}
\end{figure*}

\begin{figure*}
    \centering
    \includegraphics[scale=0.4]{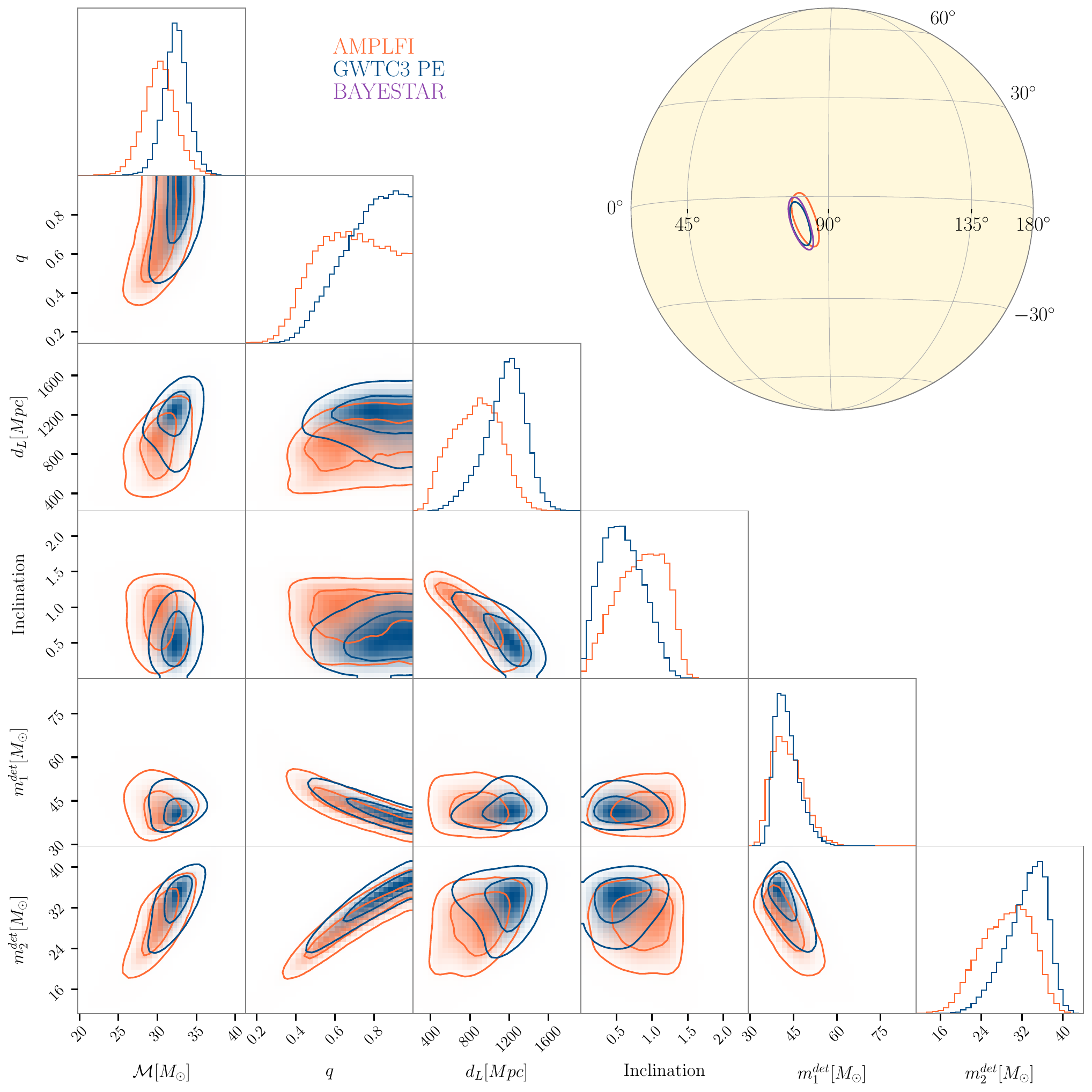}
    \caption{Posterior comparison between AMPLFI and GWTC-3 result for GW200311\_115853. Hanford, Livingston and Virgo data is analyzed. The BAYESTAR sky map produced in low-latency is plotted in purple. Sky map contours correspond to 90\% confidence intervals.}
    \label{fig:GW200311_115853-hlv}
\end{figure*}

\begin{figure*}
    \centering
    \includegraphics[scale=0.4]{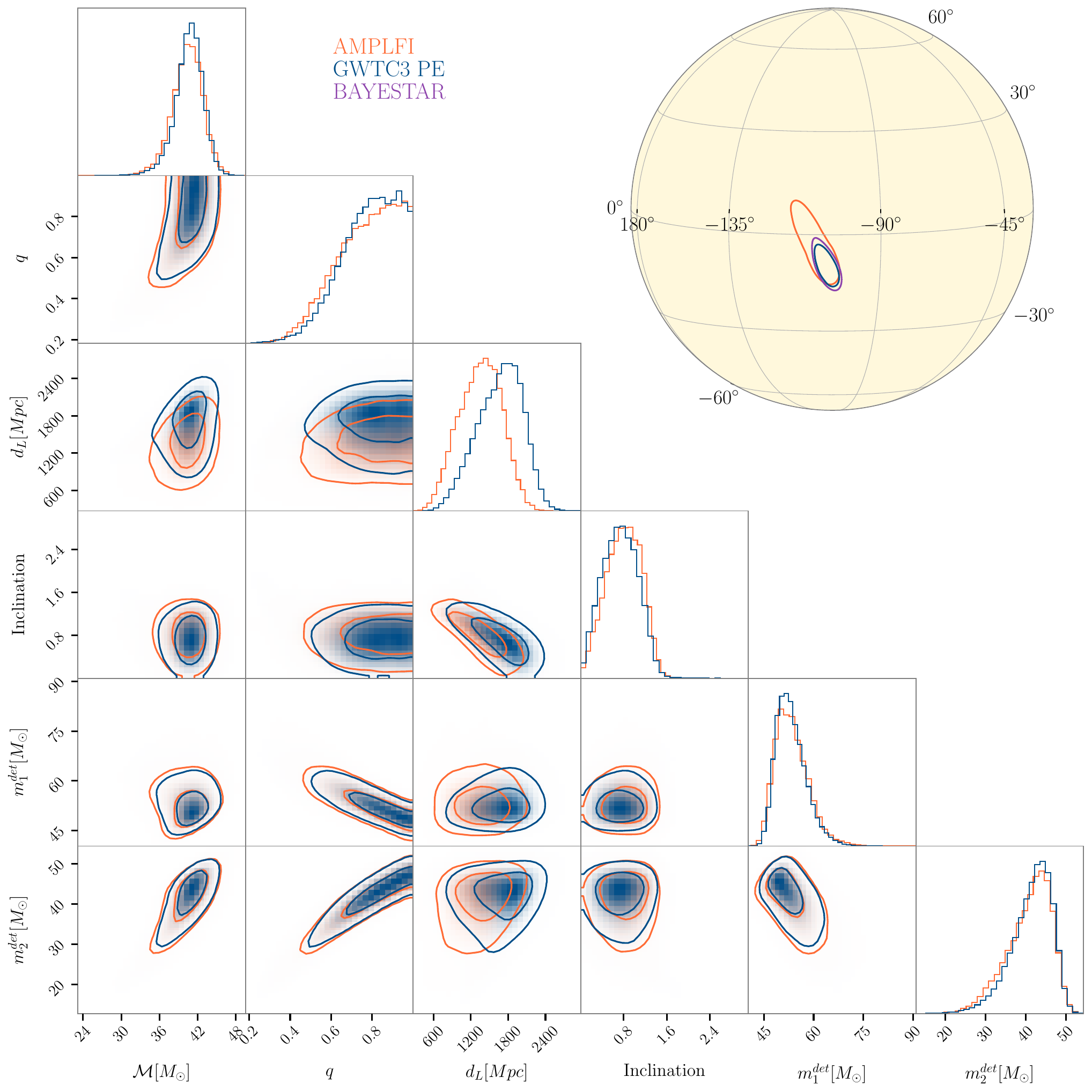}
    \caption{Posterior comparison between AMPLFI and GWTC-3 result for GW200224\_222234. Hanford, Livingston and Virgo data is analyzed.The BAYESTAR sky map produced in low-latency is plotted in purple. Sky map contours correspond to 90\% confidence intervals.}
    \label{fig:GW200224_222234-hlv}
\end{figure*}

\begin{figure*}
    \centering
    \includegraphics[scale=0.4]{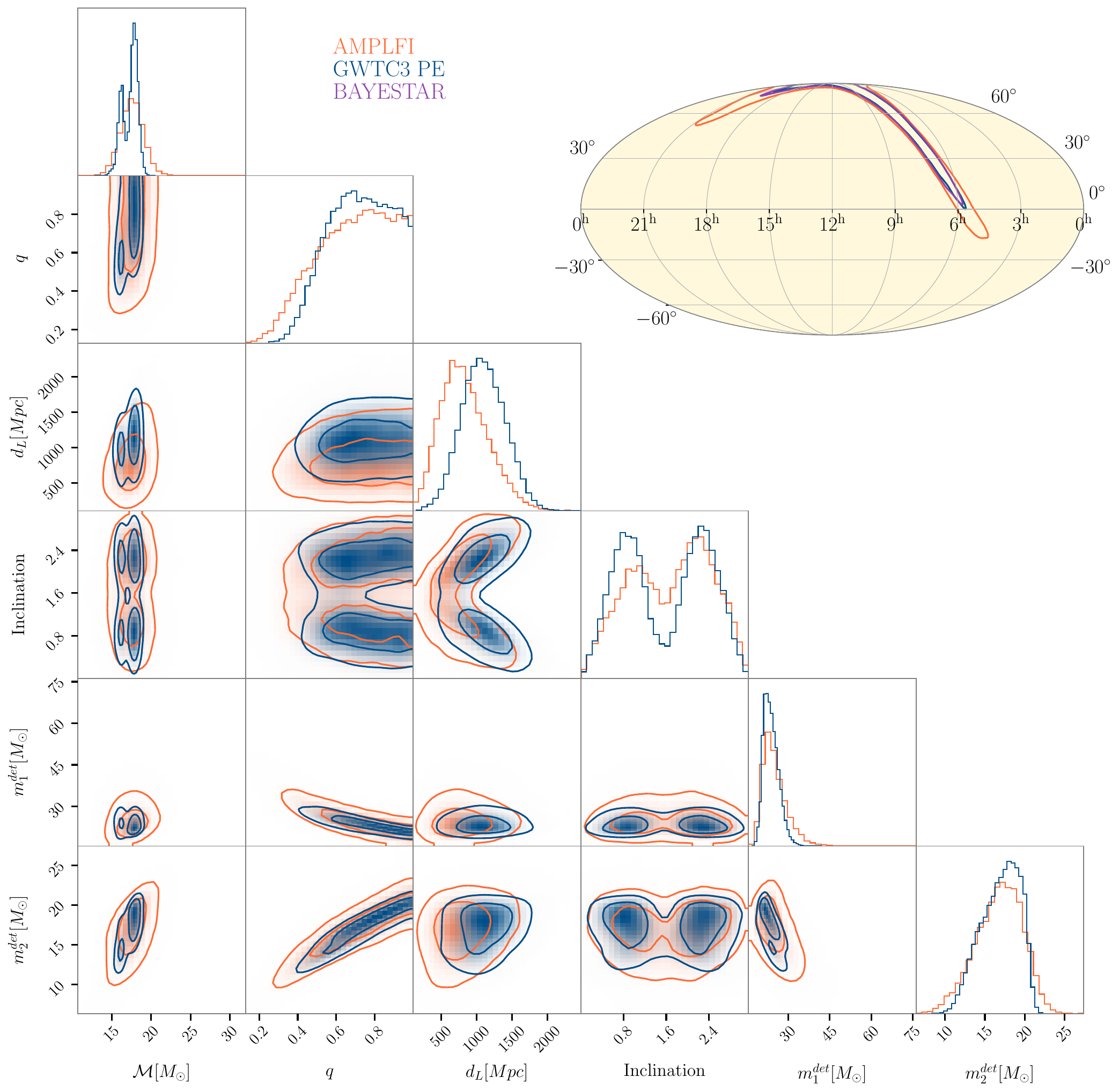}
    \caption{Posterior comparison between AMPLFI and GWTC-3 result for GW200225\_060421. Hanford and Livingston data is analyzed. The BAYESTAR sky map produced in low-latency is plotted in purple. Sky map contours correspond to 90\% confidence intervals.}
    \label{fig:GW200225_060421-hl}
\end{figure*}

\begin{figure*}
    \centering
    \includegraphics[scale=0.4]{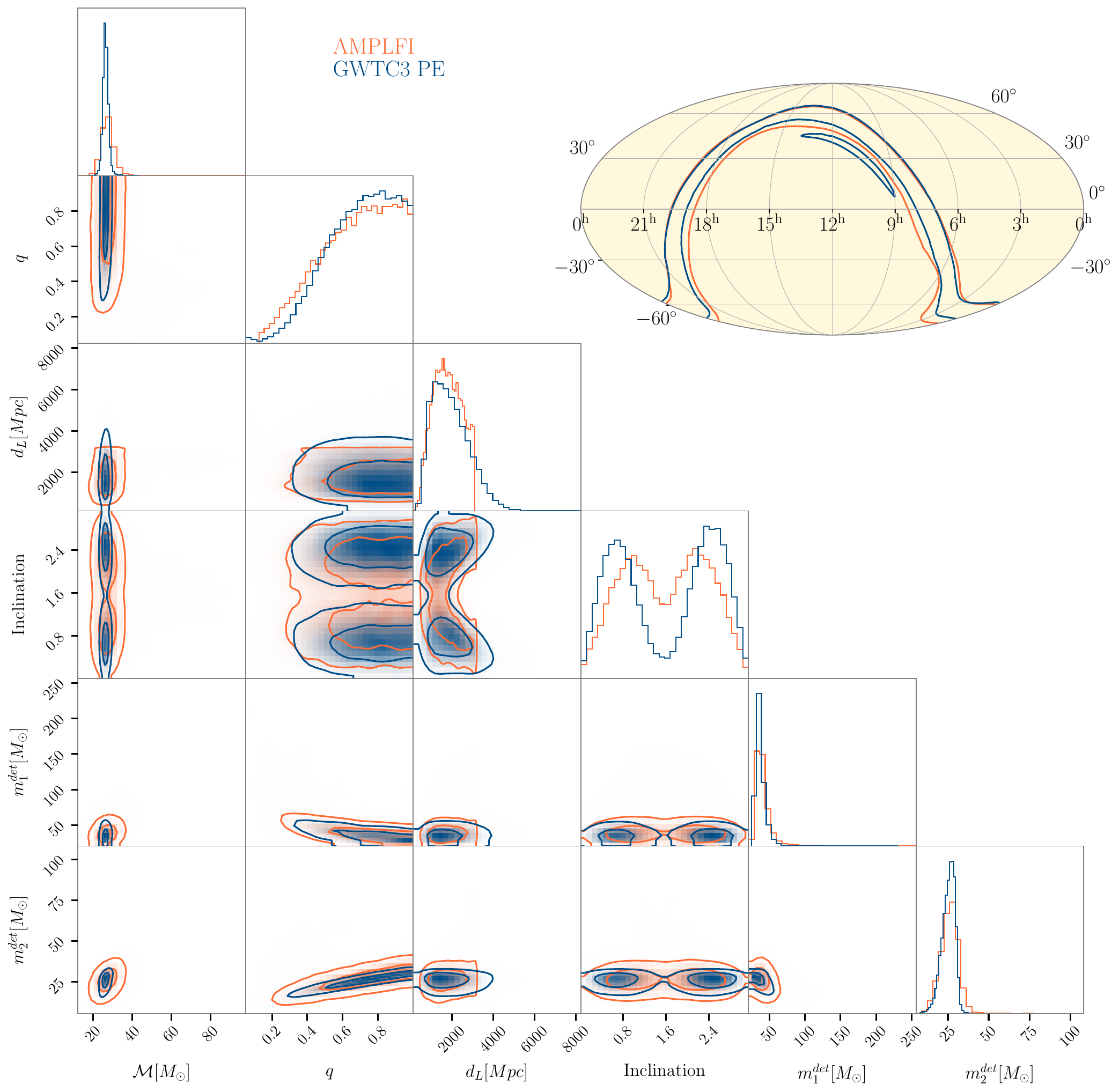}
    \caption{Posterior comparison between AMPLFI and GWTC-3 result for GW191204\_110529. Hanford and Livingston data is analyzed. Sky map contours correspond to 90\% confidence intervals.}
    \label{fig:GW191204_110529-hl}
\end{figure*}

\begin{figure*}
    \centering
    \includegraphics[scale=0.4]{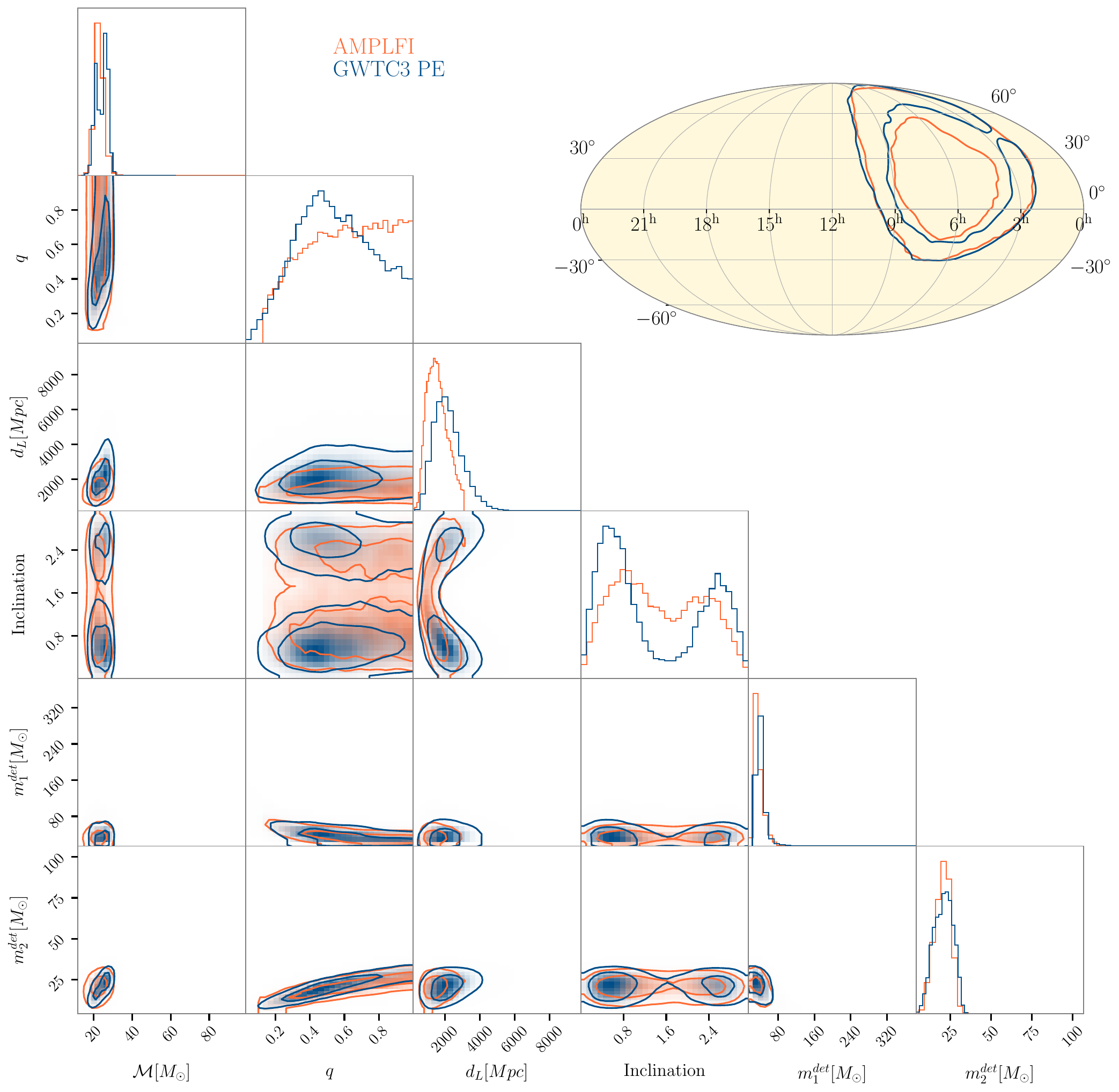}
    \caption{Posterior comparison between AMPLFI and GWTC-3 result for GW200306\_093714. Hanford and Livingston data is analyzed. Sky map contours correspond to 90\% confidence intervals.}
    \label{fig:GW200306_093714-hl}
\end{figure*}


\begin{figure*}
    \centering
    \includegraphics[scale=0.4]{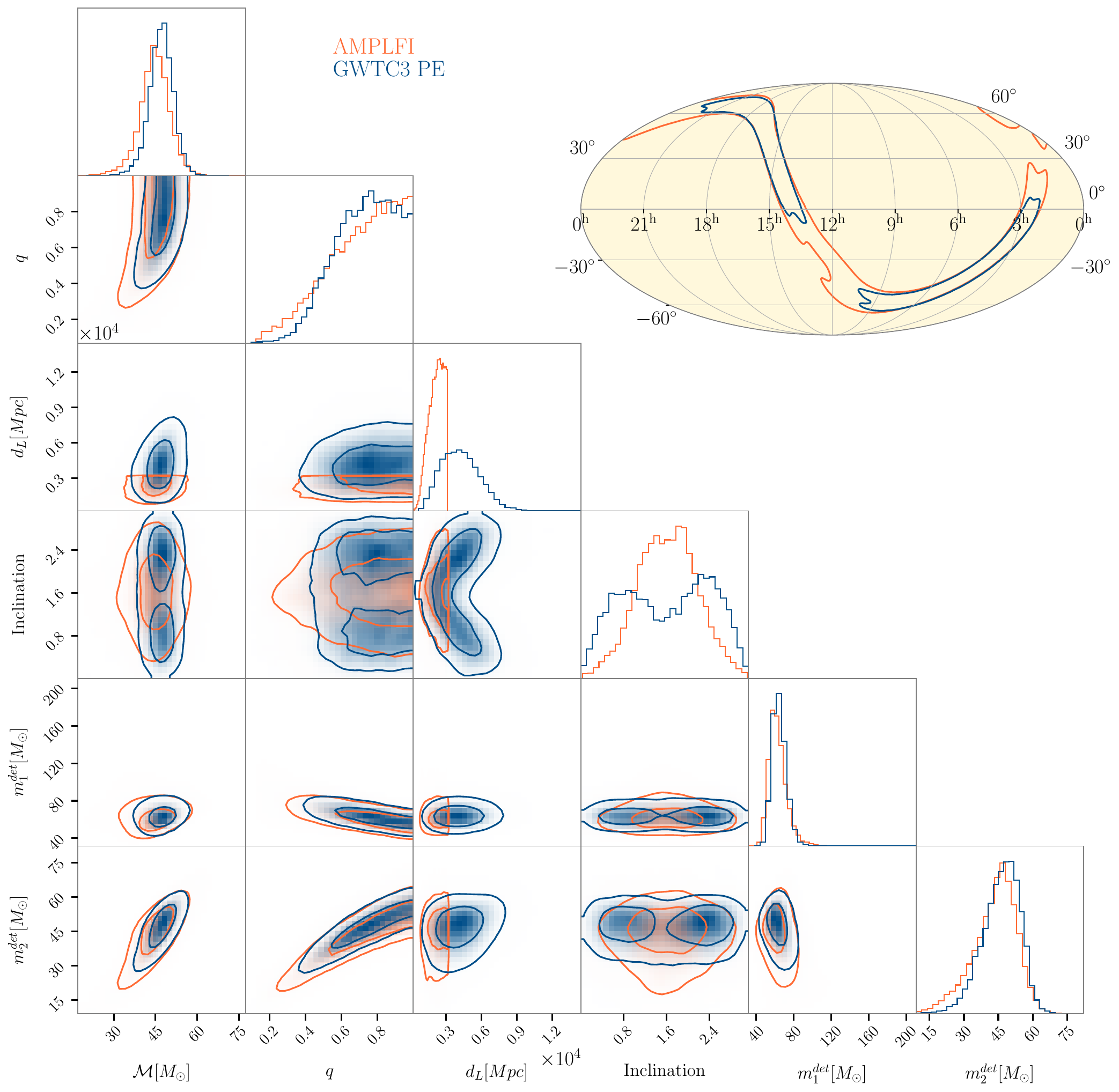}
    \caption{Posterior comparison between AMPLFI and GWTC-3 result for GW200220\_124850. Hanford and Livingston data is analyzed. This illustrates the scenario where the distance posterior has support well beyond AMPLFI's training prior. Still, AMPLFI is able to recover source parameters and localizations consistent with the GWTC-3  result.}
    \label{fig:GW200220_124850-dist-railing}
\end{figure*}

\begin{figure*}
    \centering
    \includegraphics[scale=0.4]{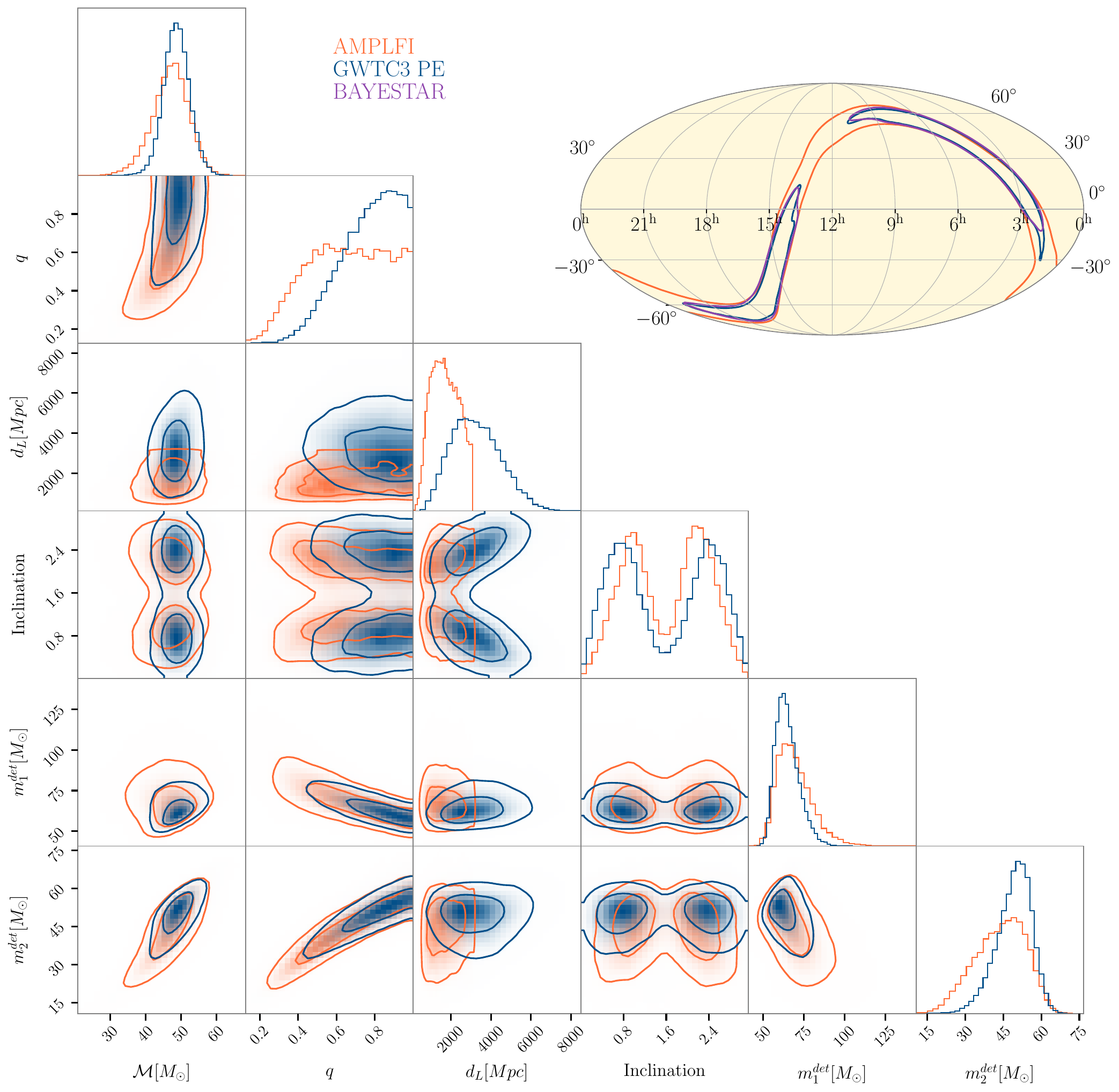}
    \caption{Posterior comparison between AMPLFI and GWTC-3 result for GW200128\_022011. Hanford and Livingston data is analyzed. The BAYESTAR sky map produced in low-latency is plotted in purple. This illustrates the scenario where the distance posterior has support well beyond AMPLFI's training prior. Still, AMPLFI is able to recover source parameters and localizations consistent with the GWTC-3  result.}
    \label{fig:GW200128_022011-dist-railing}
\end{figure*}

\begin{figure*}
    \centering
    \includegraphics[scale=0.4]{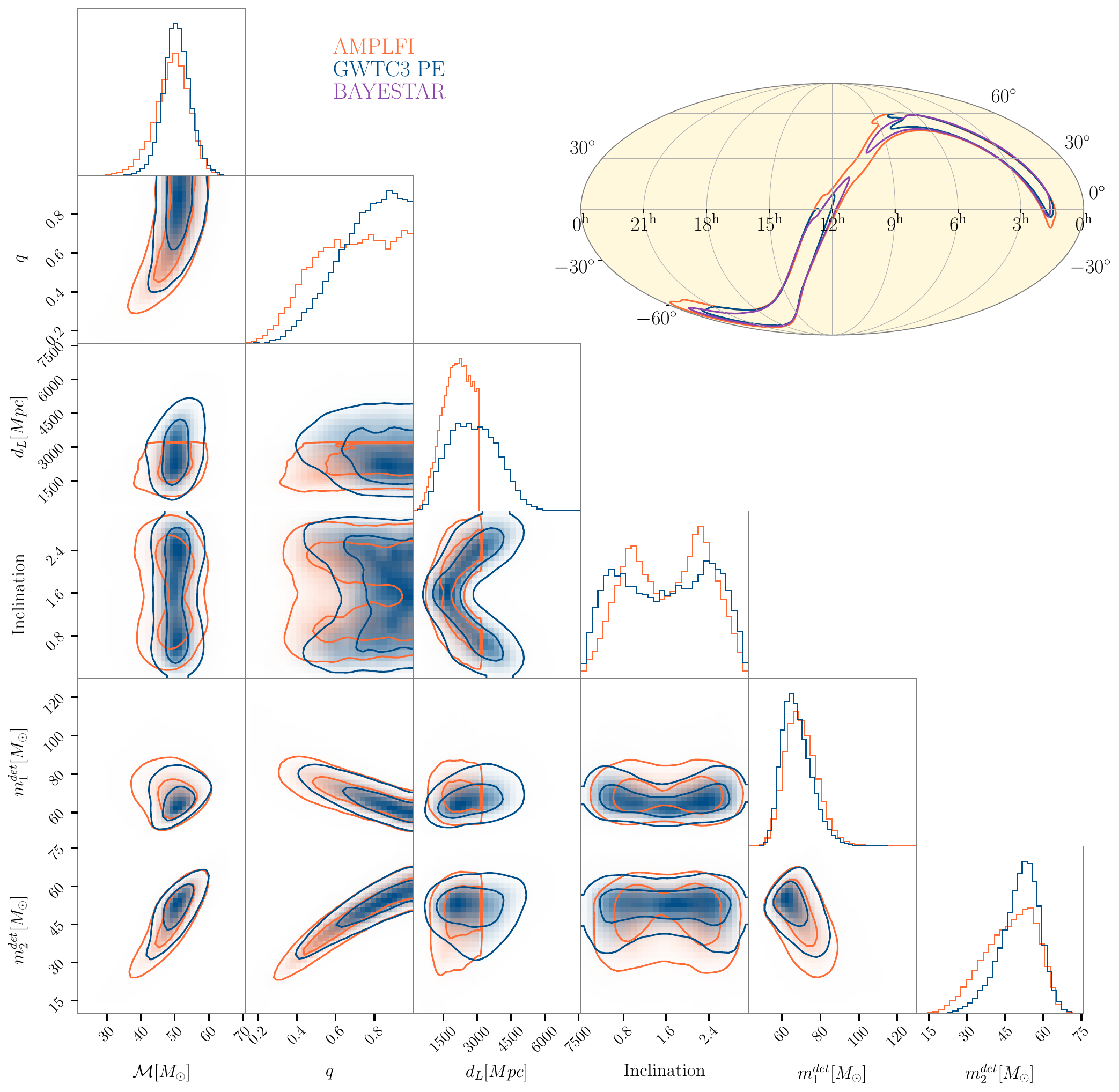}
    \caption{Posterior comparison between AMPLFI and GWTC-3 result for GW191222\_033537. Hanford and Livingston data is analyzed. The BAYESTAR sky map produced in low-latency is plotted in purple. This illustrates the scenario where the distance posterior has support well beyond AMPLFI's training prior. Still, AMPLFI is able to recover source parameters and localizations consistent with the GWTC-3 result.}
    \label{fig:GW191222_033537-dist-railing}
    
\end{figure*}

\end{document}